\documentclass[aps,prl,twocolumn,showpacs,superscriptaddress,groupedaddress]{revtex4}  
\usepackage{graphicx}  
\usepackage{dcolumn}   
\usepackage{bm}        
\usepackage{amssymb}   
\usepackage{amsmath}
\usepackage{xcolor}
\usepackage{booktabs,multirow,array,makecell}
\begin{document}

\widetext

\title{Evolution of the phenomenologically determined collective potential along the chain of Zr isotopes}

\author{E.V. Mardyban}%
\author{E.A. Kolganova}
\address{Joint Institute for Nuclear Research, 141980 Dubna, Moscow region, Russia}
\address{Dubna State University, 141982 Dubna, Moscow Region, Russia}
\author{T.M. Shneidman}
\address{Joint Institute for Nuclear Research, 141980 Dubna, Moscow region, Russia}
\address{Kazan Federal University, Kazan 420008, Russia}
\author{R.V. Jolos}
\address{Joint Institute for Nuclear Research, 141980 Dubna, Moscow region, Russia}
\address{Dubna State University, 141982 Dubna, Moscow Region, Russia}

\begin{abstract}
\begin{description}
\item[Background:] The properties of the collective low-lying states of Zr isotopes which include excitation energies and E2 reduced transition probabilities indicate that some of these states are mainly spherical and the other are mainly deformed ones. A consideration of these data in the  framework of the Geometrical Collective Model with $\beta$, $\gamma$, and rotational degrees of freedom is necessary for $^{92-102}$Zr.
\item[Purpose:] To investigate the properties of the low-lying collective states of $^{92-102}$Zr  based on the five-dimensional Geometrical Quadrupole Collective Model, to obtain the collective potentials for the chains of Zr isotopes and to investigate their evolution with increase of the number of neutrons.
\item[Method:] The quadrupole-collective Bohr Hamiltonian depending on both $\beta$ and $\gamma$ shape variables with a potential having spherical and deformed minima,  is applied. The  relative depth of two minima,  height and width of the barrier, rigidity of the potential  near both minima are determined so as to achieve the best possible description of the observed properties of the low-lying collective quadrupole states of $^{92-102}$Zr.
\item[Results:]  Satisfactory agreement with the experimental data on the excitation energies and the E2  reduced transition probabilities is obtained. The evolution of the collective potential with increase of $A$ is described and the distributions of the wave functions of the collective states  in $\beta-\gamma$ plane are found.
\item[Conclusion:] It is shown that the low-energy structure of $^{92-102}$Zr can be described in a satisfactory way within the Geometrical Collective Model with the Bohr Hamiltonian.  The  $\beta$-dependence of the potential energy is fixed to describe the experimental data in a best possible way. The resulting potential evolves with $A$ increase from having only one spherical minimum in $^{92}$Zr, through the potentials having both spherical and deformed minima, to the potential with one deformed minimum in $^{102}$Zr. A $\beta$-dependence of the wave functions is presented in a set of figures illustrating their distribution over $\beta$.

\end{description}
\end{abstract}

\maketitle

\section{Introduction}

Zr isotopes are of particular interest for investigation of nuclear structure, especially
because they are characterized by a dramatic transition from spherical to deformed shape at low excitation energies
\cite{Heyde1,Garcia}. Many spectroscopic data have been accumulated for these nuclei
\cite{Kawade,Ebert,Mach1,Mach2,Lhersonneau,Urban,Wu,Hager,Goodina,Chakrabarty,Browne,Kremer,Ansari,Singh,Witt,Werner} containing information on excitation energies and electromagnetic transition probabilities. Various theoretical approaches have been used to study shape phase transitions in
these nuclei \cite{Garcia,Federman1,Federman2,Bonche,Kumar,Galeriu,Moeler,Skalski,Lalazissis,Holt,Xu,Sieja,Togashi,Skalski1,Rodriguez,Delaroche,Reinhardt,Liu,
Petrovici,Ozen,Nomura,Xiang,Otsuka,Mei,Verma,Gregor,Miyahara,Stone,Niksic2,Vitturi,Sugita,Lalkovski,Isacker,Boyukata,Fortune,Garcia2,Gavrielov1,Gavrielov2,Garcia-Ramos,Sazonov,Mardyban,Niksic1,Fortune1}. Theoretical studies of Zr isotopes can be divided into two groups.
The  first group was aimed to clarify  the mechanism of a sharp transition from the spherical shape  to the deformed one. These investigations were based either on the nuclear Shell Model, or on the Hartree-Fock-Bogoliubov approach and on the Energy Density Functional \cite{Federman1,Federman2,Bonche,Kumar,Galeriu,Moeler,Skalski,Lalazissis,Holt,Xu,Sieja,Togashi,Skalski1,Rodriguez,Delaroche,Reinhardt,Liu,
Petrovici,Ozen,Nomura,Xiang,Otsuka,Mei,Verma,Gregor,Miyahara,Stone,Niksic2}. The second group includes studies based mainly on the phenomenological models \cite{Sugita,Lalkovski,Isacker,Boyukata,Fortune,Garcia2,Gavrielov1,Gavrielov2,Garcia-Ramos,Sazonov,Mardyban}.
Their aim was to achieve the best possible description of experimental data on Zr isotopes, fixing the model parameters for this purpose. This approach  allows  to come to a qualitative characteristic of the low-lying excited states, using the terminology of the nuclear collective model.

In this work, the consideration of the low-lying states of  $^{92-102}$Zr is based on the five-dimensional Bohr Hamiltonian. The aim of this work is to determine the collective potentials for the $^{92-102}$Zr isotopes fixing their  $\beta$-dependence so as to obtain the better possible description of the experimental data and investigate the evolution of the collective potential with increase of the number of neutrons. The other aim is  to get a distribution of the wave functions of the low-lying collective states in $\beta-\gamma$ plane.

This consideration is not free from restrictions. A certain dependence of the potential on a variable describing non-axial deformation is assumed and restrictions on the inertia tensor are imposed.

\section{Hamiltonian}

In general case the quadrupole collective Bohr Hamiltonian can be presented as (see \cite{Niksic1})
\begin{widetext}
\begin{eqnarray}
\label{general}
H&=& -\frac{\hbar^2}{2 \sqrt{\omega r}} \left (\frac{1}{\beta^4} \left [ \frac{\partial}{\partial \beta}\sqrt{\frac{r}{\omega}}\beta^4 B_{\gamma \gamma} \frac{\partial}{\partial \beta}\right.\right.\nonumber-\left.\left.\frac{\partial}{\partial \beta}\sqrt{\frac{r}{\omega}}\beta^3 B_{\beta \gamma}\frac{\partial}{\partial \gamma}\right ]\right. \nonumber \\
&+&\left. \frac{1}{\beta \sin{3\gamma}} \left [ -\frac{\partial}{\partial \gamma} \sqrt{\frac{r}{\omega}}\sin{3\gamma}B_{\beta \gamma}\frac{\partial}{\partial \beta}\right.\right.
+\left.\left. \frac{1}{\beta}\frac{\partial}{\partial \gamma}\sqrt{\frac{r}{\omega}}\sin{3\gamma}B_{\beta \beta}\frac{\partial}{\partial \gamma}\right ] \right ) +\frac{1}{2} \sum_{k=1}^3 \frac{\hat J^2_k}{\Im_k(\beta)}+V(\beta, \gamma).
\end{eqnarray}
\end{widetext}
Here $\omega = B_{\beta \beta}B_{\gamma \gamma}-B^2_{\beta \gamma}$ is the determinant of the vibrational part of the inertia tensor
\begin{eqnarray}
\label{Bvib}
B_{vib} =
\begin{pmatrix}
B_{\beta \beta} & \beta B_{\beta \gamma} \\
\beta B_{\beta \gamma}& \beta^2 B_{\gamma \gamma}
\end{pmatrix},
\end{eqnarray}
and $r=B_1 B_2 B_3$, where $B_k$ (k=1,2,3) are presented in the expression for the components of the moment of inertia $\Im_k$ determined with respect to the body-fixed axes as
\begin{eqnarray}
\label{moment_of_inertia}
\Im_k=4 B_k(\beta)\beta^2 \sin^2{\left (\gamma-\frac{2\pi k}{3}\right )}.
\end{eqnarray}
The components of the angular momentum in the body-fixed frame are denoted as $\hat J_k$ and can be expressed in terms of the  Euler angles. The potential energy is denoted as $V(\beta,\gamma)$. The Hamiltonian (\ref{general}) is a general case of the conventional Bohr Hamiltonian \cite{Bohr1952} allowing the non-zero value of $B_{\beta\gamma}$.

To simplify consideration, we make below the following  assumptions for the inertia coefficients:
\begin{eqnarray}
\label{assumptions}
&& B_{\beta \beta}=B_{\gamma \gamma}=B_0, \hspace{10pt}  B_{\beta \gamma}=0,\nonumber \\
&& B_1(\beta)=B_2(\beta)=B_3(\beta)=b_{rot}(\beta) B_0,
\end{eqnarray}
where $B_0$ is the parameter scaling vibrational and rotational inertia coefficients. With respect to the value of $b_{rot}$, it is known \cite{Jolos1,Jolos2} that in the deformed nuclei $b_{rot}$ is less than one. In the region of a small values of $\beta$ we put $b_{rot}=1$ (see discussion in \cite{Mardyban}). Since a general feature of the collective potential used below is the presence of two minima, spherical and deformed, separated by the barrier at $\beta=\beta_m$ we set $b_{rot}$ as following
\begin{eqnarray}
\label{brot}
b_{rot}=
        \begin{cases}
            1 & \text{if $\beta \le \beta_m$, } \\
            b_{def}<1 & \text{if $\beta > \beta_m$.}
        \end{cases}
\end{eqnarray}
The change from the spherical to deformed value of $b_{rot}$ occurs at $\beta =\beta_m$ which is taken around the maximum of the barrier separating spherical and deformed potential wells. Our calculations show that the small variations of $\beta_m$ does not affect  qualitatively the results of the calculations.

With these assumptions about inertia tensor the Hamiltonian (\ref{general}) takes the form:
\begin{eqnarray}
\label{Hamiltonian_simplified_1}
\hat H &=& -\frac{\hbar^2}{2 B_0} \left (\frac{1}{b_{rot}^{3/2}}\frac{1}{\beta^4} \frac{\partial}{\partial \beta}\beta^4 b_{rot}^{3/2} \frac{\partial}{\partial \beta} \right. \\
&+& \left.\frac{1}{\beta^2 \sin{3\gamma}} \frac{\partial}{\partial \gamma}\sin{3\gamma}\frac{\partial}{\partial \gamma}\right )
+ \frac{1}{2} \sum_{k=1}^3 \frac{\hat J^2_k}{\Im_k(\beta)}+V(\beta, \gamma). \nonumber
\end{eqnarray}

We expect that the wave functions of the lowest states are localized in the minima while the weight of the wave functions inside the barrier region is strongly suppressed. In this case, we can neglect a derivative of $b_{rot}$ over $\beta$ which is presented, in principle, in the kinetic part of the Hamiltonian (\ref{Hamiltonian_simplified_1}), as it gives the non-zero contribution to the matrix elements of the Hamiltonian only in the barrier region where the wave functions are close to zero. Thus, we obtain finally the following model Hamiltonian:
\begin{eqnarray}
\label{Hamiltonian_simplified_2}
H&=& -\frac{\hbar^2}{2 B_0} \left (\frac{1}{\beta^4} \frac{\partial}{\partial \beta}\beta^4 \frac{\partial}{\partial \beta}
+\frac{1}{\beta^2 \sin{3\gamma}} \frac{\partial}{\partial \gamma}\sin{3\gamma}\frac{\partial}{\partial \gamma}\right. \nonumber \\
&+& \left. \sum_{k=1}^3 \frac{\hat J^2_k}{4 b_{rot} \beta^2 \sin{\left (\gamma - \frac{2 \pi k}{3}\right )}}\right )+V(\beta, \gamma),
\end{eqnarray}

The aim of the present work is to determine the potential energy in the Bohr Hamiltonian for nuclei belonging to the chain of the Zr isotopes so as to achieve the best possible agreement with the experimental data in excitation energies and E2 transition probabilities for the low lying states. This  give us the possibility to explore the evolution of the collective potential with increase of $A$ along the chain of Zr isotopes.  The potential energy $V(\beta,\gamma)$ in \eqref{Hamiltonian_simplified_2} is chosen in the form
\begin{eqnarray}
V(\beta,\gamma)=U(\beta) +C_\gamma \beta^3(1-\cos{3\gamma}).
\label{potential_energy}
\end{eqnarray}

Thus, we fix the dependence of $V$ on $\gamma$, eliminating potentials with minimum corresponding to the non-axial shapes. The form of $V(\beta)$ and the parameter $C_\gamma$, which determines the stiffness of the potential with respect to $\gamma$ oscillations at the deformed minimum, are fitted to reproduce the experiment data.

To describe the shape of $V(\beta)$ for $^{96}$Zr we have defined in \cite{Sazonov}  several points fixing the positions of the spherical and deformed minima, the rigidity of the potential near its minima, and the height and width of the barrier separating two minima. Then we vary the positions of the selected points in order to get a satisfactory description of the experimental data. The number of points was taken to be 16 to provide a smooth change of the potential with $\beta$. However, not all the points are of the same physical importance. In principle, the number of points can be
 minimized as, obviously, the only relative depths of the minima and the height and width of the barrier leads to physically meaningful changes. For this reason, in the present paper we approximate the potential by the three parabolas: two to describe spherical and deformed minima, and the inverted parabola to describe the barrier separating two minima. The parameters of the parabolas have been fixed so as it is described above.

 To solve the eigenvalue problem with the Hamiltonian (\ref{Hamiltonian_simplified_2})  we expand the eigenfunctions in terms of a complete set of basis functions
that depends on the deformation variables $\beta$ and $\gamma$, and the Euler angles. These functions are well known and their construction is described in the literature. See, for instance, \cite{Bes,Dussel} and references below. For completeness of presentation we give some details here.

For each value of angular momentum $I$, the basis functions are written as
\begin{eqnarray}
\label{basis1}
\Psi_{IM}^{n_\beta v \alpha}=R^{(n_\beta, v)}(\beta)\Upsilon_{v \alpha I M}(\gamma, \Omega),
\end{eqnarray}
where $\Upsilon_{v \alpha I M}$ are the SO(5)$ \supset$ SO(3) spherical harmonics, which are the eigenfunctions of the operator $\hat \Lambda^2$:
\begin{eqnarray}
&\hat \Lambda^2&\Upsilon_{v \alpha I M} \nonumber\\
&=&\left[-\frac{1}{\sin{3\gamma}} \frac{\partial}{\partial \gamma} \sin{3\gamma} \frac{\partial}{\partial \gamma} +
 \frac{1}{4} \sum_{k} \frac{\hat J_k^2}{\sin^2{(\gamma -\frac{2\pi k}{3})}} \right ]\Upsilon_{v \alpha I M}  \nonumber\\
&=&v (v+3)\Upsilon_{v \alpha I M}.
\label{laplacian}
\end{eqnarray}
In addition to the angular momentum $I$ and its projection $M$, each function $\Upsilon_{v \alpha I M}$ is labeled by the SO(5) seniority quantum number $v$ and a multiplicity index $\alpha$,
 which is required for $v \ge 6$.

The $\Upsilon_{v \alpha I M}$ can be explicitly constructed as a sum over the functions with explicit value of the projection $K$ of the angular momentum on the intrinsic axis
\cite{Rowe2004,Caprio2009}
\begin{eqnarray}
\Upsilon_{v \alpha I M}(\gamma, \Omega)=\sum_{K=0, even}^{I}F_{v \alpha I,K}(\gamma)\xi_{KM}^{I}(\Omega),
\end{eqnarray}
where
\begin{eqnarray}
\label{eq11}
\xi_{KM}^{I}(\Omega)=\frac{1}{\sqrt{2(1+\delta_{K0})}}\left [D^I_{M\  K}(\Omega)\right.\nonumber\\
\left.
+(-1)^I D^I_{M\  -K}(\Omega)\right ]
\end{eqnarray}
 and the $F_{v \alpha I,K}(\gamma)$ are polynomials constructed from the trigonometrical functions of $\gamma$ \cite{Rowe-Wood}. This construction is used in the present work.

The basis wave functions $R^{(n_\beta, v)}$  are chosen as the eigenfunctions of the harmonic oscillator Hamiltonian in $\beta$:
\begin{eqnarray}
h_{h.o.}=\frac{1}{2}\left(-\frac{1}{\beta^4}\frac{\partial}{\partial\beta}\beta^4\frac{\partial}{\partial\beta}+\frac{v(v+3)}{\beta^2}+\frac{\beta^2}{\beta_0^4}\right).
\label{oscillator}
\end{eqnarray}

The eigenfunctions of $h_{h.o.}$ have the following analytical form
\begin{eqnarray}
\label{sol_1}
R_{n_\beta, v}(\beta)=N_\beta \left (\frac{\beta}{\beta_0}\right) ^v L_{n_\beta}^{v+3/2}\left (\frac{\beta^2}{\beta_0^2} \right ) \exp{\left ( -\frac{\beta^2}{2\beta_0^2} \right )},
\label{basis_beta}
\end{eqnarray}
where $\beta_0$ is an oscillator length and the normalization constant  $N_\beta$ is given as:
\begin{equation}
\label{N_const}
N_\beta=\sqrt{ \frac{2n_\beta!}{\Gamma(n_\beta+v+5/2)} }.
\end{equation}

The basis  functions $R_{n_\beta, v}$ are completely specified by the choice of the  oscillator length $\beta_0$. Our calculations have shown that the
 fastest  convergence of the results is obtained when $\beta_0$ is chosen to be equal to the value at the region of the barrier separating two minima, so that the oscillator potential coincides with the potential $V(\beta)$ at the top of the barrier.  For such a choice of $\beta_0$, $(n_\beta)_{max} = 30$ is enough to provide a convergence. The basis of harmonics $\Upsilon_{v \alpha I M}(\gamma, \Omega)$  is truncated to some maximum  seniority $v_{max}$. As shown in \cite{Caprio2011}, taking $v_{max}=50$ is sufficient to provide a convergence of the calculation.

 The Hamiltonian eigenfunctions $\Psi_{InM}$, where  $n$ is a multiplicity index, are obtained in calculations as a series expansions in the basic functions
(\ref{basis1}). For discussions below it is convenient to introduce the following quantities, namely,\\
- the distributions of the squares of the wave functions in the $\beta-\gamma$ plane:
 \begin{eqnarray}
\label{distribution1}
 \beta^4 \sin{3 \gamma}\int d\Omega  |\Psi_{InM}|^2,
\end{eqnarray}
- the one-dimensional probability distributions over $\beta$ determined by integration of
$|\Psi_{InM}|^2$ over $\gamma$ and Euler angles:
\begin{eqnarray}
\label{distribution2}
\beta^4 \int_{0}^{\pi/3}\sin{3 \gamma}d\gamma \int d\Omega  |\Psi_{InM}|^2.
\end{eqnarray}

\begin{table*}[tbh]
\centering
\setlength\aboverulesep{0pt}\setlength\belowrulesep{0pt}
\setcellgapes{3pt}\makegapedcells
\caption{Comparison of the experimental and calculated energies of the low-lying collective states of the even-even $^{92-102}$Zr isotopes.
The observed $2^+_2$ state in $^{92}$Zr has a noncollective nature and is presented for the completeness only. For this reason, the calculated excitation energy
of this state is not shown in this table.}
\begin{tabular}{l|cc|cc|cc|cc|cc|cc}
\hline
\hline
A	&	\multicolumn{2}{c|}{92}		&	\multicolumn{2}{c|}{94}		&	\multicolumn{2}{c|}{96}		&	\multicolumn{2}{c|}{98}		&	\multicolumn{2}{c|}{100}		&	 \multicolumn{2}{c}{102}		 \\
\hline
State (MeV)&	exp	&	calc	&	exp	&	calc	&	exp	&	calc	&	exp	&	calc	&	exp	&	calc	&	exp	&	calc	\\
\hline
$E(0_2^+)$	&	1.38	&	1.35	&	1.30	&	1.30	&	1.58	&	1.58	&	0.85	&	0.86	&	0.33	&	0.33	&	0.90	&	0.90	\\
$E(0_3^+)$	&	2.90	&	2.10	&		&	1.57	&	2.70	&	2.40	&	1.44	&	1.63	&	0.83	&	0.80	&		&	1.38	\\
$E(2_1^+)$	&	0.93	&	0.91	&	0.92	&	0.95	&	1.75	&	1.75	&	1.22	&	1.18	&	0.21	&	0.19	&	0.15	&	0.15	\\
$E(2_2^+)$	&	1.85	&	--	&	1.67	&	1.61	&	2.23	&	2.00	&	1.59	&	1.59	&	0.83	&	0.88	&	1.04	&	1.02	\\
$E(2_3^+)$	&	2.07	&	1.77	&	2.15	&	1.92	&	2.67	&	2.79	&	1.75	&	2.00	&	1.20	&	1.22	&	1.21	&	1.21	\\
$E(3_1^+)$	&	2.91	&	2.91	&	2.51	&	3.19	&	2.44	&	3.74	&		&	2.80	&	1.40	&	1.75	&	1.24	&	1.32	\\
$E(4_1^+)$	&	1.50	&	1.88	&	1.47	&	1.98	&	2.86	&	2.69	&	1.84	&	1.86	&	0.56	&	0.56	&	0.48	&	0.47	\\
$E(4_2^+)$	&	2.40	&	2.56	&	2.33	&	2.07	&	3.08	&	3.34	&	2.05	&	2.53	&	1.41	&	1.43	&	1.39	&	1.39	\\
$E(4_3^+)$	&	2.87	&	2.90	&	2.86	&	2.77	&	3.18	&	3.68	&	2.28	&	2.95	&	1.86	&	1.89	&	1.54	&	1.68	\\
\hline
\hline
\end{tabular}
\end{table*}

\begin{table*}[tbh]
\centering
\setlength\aboverulesep{0pt}\setlength\belowrulesep{0pt}
\setcellgapes{3pt}\makegapedcells
\caption{Comparison of the experimental and calculated $E2$ reduced transition probabilities between the low-lying states of the even-even $^{92-102}$Zr isotopes.
The observed $2^+_2$ state has a noncollective nature and is presented for the completeness only. For this reason, the calculated values of the E2 reduced
transition probabilities from and to this state are not shown in the table.}
\begin{tabular}{l|cc|cc|cc|cc|cc|cc}
\hline
\hline
A	&	\multicolumn{2}{c|}{92}		&	\multicolumn{2}{c|}{94}		&	\multicolumn{2}{c|}{96}		&	\multicolumn{2}{c|}{98}		&	\multicolumn{2}{c|}{100}		&	 \multicolumn{2}{c}{102}		 \\
\hline
Transition (W.u.)&	exp	&	calc	&	exp	&	calc	&	exp	&	calc	&	exp	&	calc	&	exp	&	calc	&	exp	&	calc	\\
\hline
$B(E2;0_2\rightarrow2_1)$	&	14.4(5)	&	20.74	&	9.4(4)	&	12.94	&		&	4.90	&		&		&	67(7)	&	26.35	&		&	10.37	\\
$B(E2;2_1\rightarrow0_2)$	&		&		&		&	&		&		&	29($^{+8}_{-6}$)		&	29.02	&		&		&		&	\\
$B(E2;2_1\rightarrow0_1)$	&	6.4(6)	&	8.17	&	4.9(3)	&	8.09	&	2.3(3)	&	3.69	&	2.9(6)	&	1.84	&	77(2)	&	64.70	&	105(14)	&	53.94	\\
$B(E2;2_2\rightarrow0_1)$	&	3.7(5)	&	--	&	3.9(3)	&	0.00	&	0.26(8)	&	0.26	&		&	2.58	&		&	0.00	&		&	1.40	\\
$B(E2;2_2\rightarrow0_2)$	&		&	--	&	19(2)	&	31.59	&	36(11)	&	33.84	&		&	3.00	&		&	18.10	&		&	23.55	\\
$B(E2;2_2\rightarrow2_1)$	&	0.001($^{+25}_{-1}$)	&	--	&	0.06($^{+0.13}_{-0.06}$)	&	1.52	&	2.8($^{+1.5}_{-1.0}$)		&	3.86	&		&	12.56	&		&	 3.49	&		&	 9.79	\\
$B(E2;2_3\rightarrow0_1)$	&	$<$0.0042	&	0.02	&		&	0.01	&		&	0.01	&		&	0.25	&		&	0.53	&		&	1.92	\\
$B(E2;2_3\rightarrow2_1)$	&	$<$15	&	16.27&		&	13.76	&	50(7)	&	3.07	&		&	9.90	&		&	0.03	&		&	0.03	\\
$B(E2;3_1\rightarrow2_1)$	&	0.03($^{+9}_{-20}$)	&	0.04	&		&	0.03	&	0.1($^{+0.3}_{-0.1}$)	&	0.15	&		&	1.93	&		&	4.49	&		&	5.00	 \\
$B(E2;3_1\rightarrow2_2)$	&	2.4(14)	&	--	&		&	3.59	&		&	3.14	&		&	3.79	&		&	0.08	&		&	42.70	\\
$B(E2;4_1\rightarrow2_1)$	&	4.05(12)	&	17.65&	0.879(23)&	4.19	&	16($^{+5}_{-13}$)	&	6.86	&	42($^{+10}_{-7}$)		&	46.99	&	101.4(11)	&	104.41	&	 167($^{+30}_{-22}$)	 &	81.49	\\
$B(E2;4_1\rightarrow2_2)$	&		&	--	&		&	88.07	&	56(44)	&	55.76	&	54($^{+18}_{-16}$)		&	11.29	&		&	12.76	&		&	0.31	\\
$B(E2;4_2\rightarrow2_1)$	&	5.9(7)	&	0.002	&	13($^{+4}_{-7}$)	&	12.08	&		&	7.04	&		&	1.00	&		&	0.03	&		&	0.15	\\
$B(E2;4_2\rightarrow2_2)$	&		&	--	&	34($^{+10}_{-17}$)	&	15.22	&		&	2.29	&		&	17.41	&		&	58.67	&		&	50.37	\\
$B(E2;4_2\rightarrow4_1)$	&	2.3(13)	&	5.55	&		&	12.26	&		&	3.30	&		&	9.87	&		&	3.73	&		&	9.43	\\
$B(E2;4_3\rightarrow2_1)$	&	0.76(15)	&	0.01	&		&	0.04	&		&	0.21	&		&	0.02	&		&	0.58	&		&	0.73	\\
$B(E2;4_3\rightarrow4_1)$	&	2.7(6)	&	6.84	&		&	1.16	&		&	6.16	&		&	1.92	&		&	2.71	&		&	0.32	\\
$B(E2;4_3\rightarrow4_2)$	&	0.03($^{+99}_{-3}$)		&	8.93	&		&	4.56	&		&	0.13	&		&	8.97	&		&	11.48	&		&	6.64\\
\hline
\hline
\end{tabular}
\end{table*}
\section{Results}
The results of calculations of the excitation energies and the $E2$ reduced transition probabilities are presented in Tables I and II.
\begin{figure*}[htb]
\centering
(a)
\includegraphics
[width=0.3\textwidth]{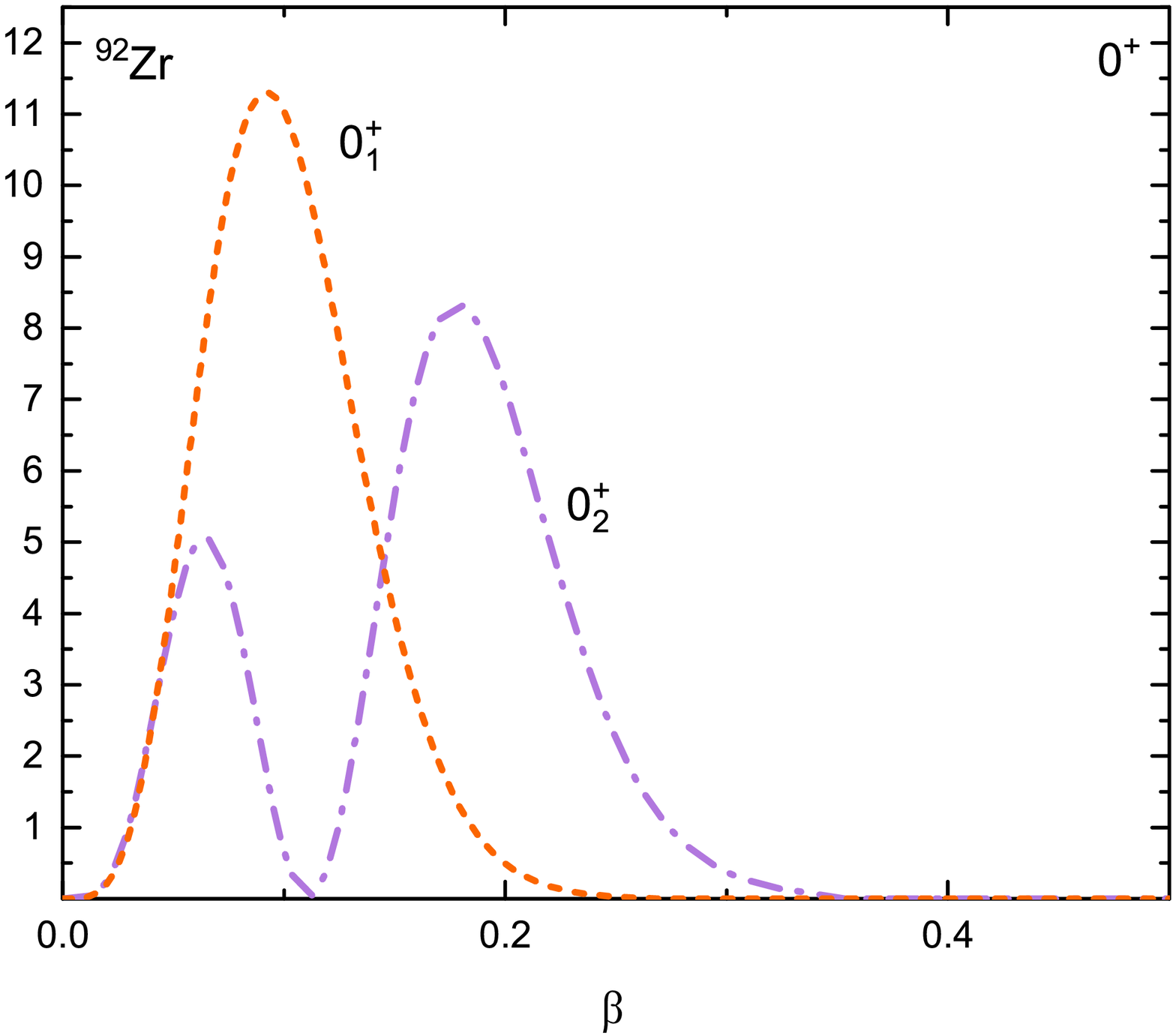}
(b)
\includegraphics
[width=0.3\textwidth]{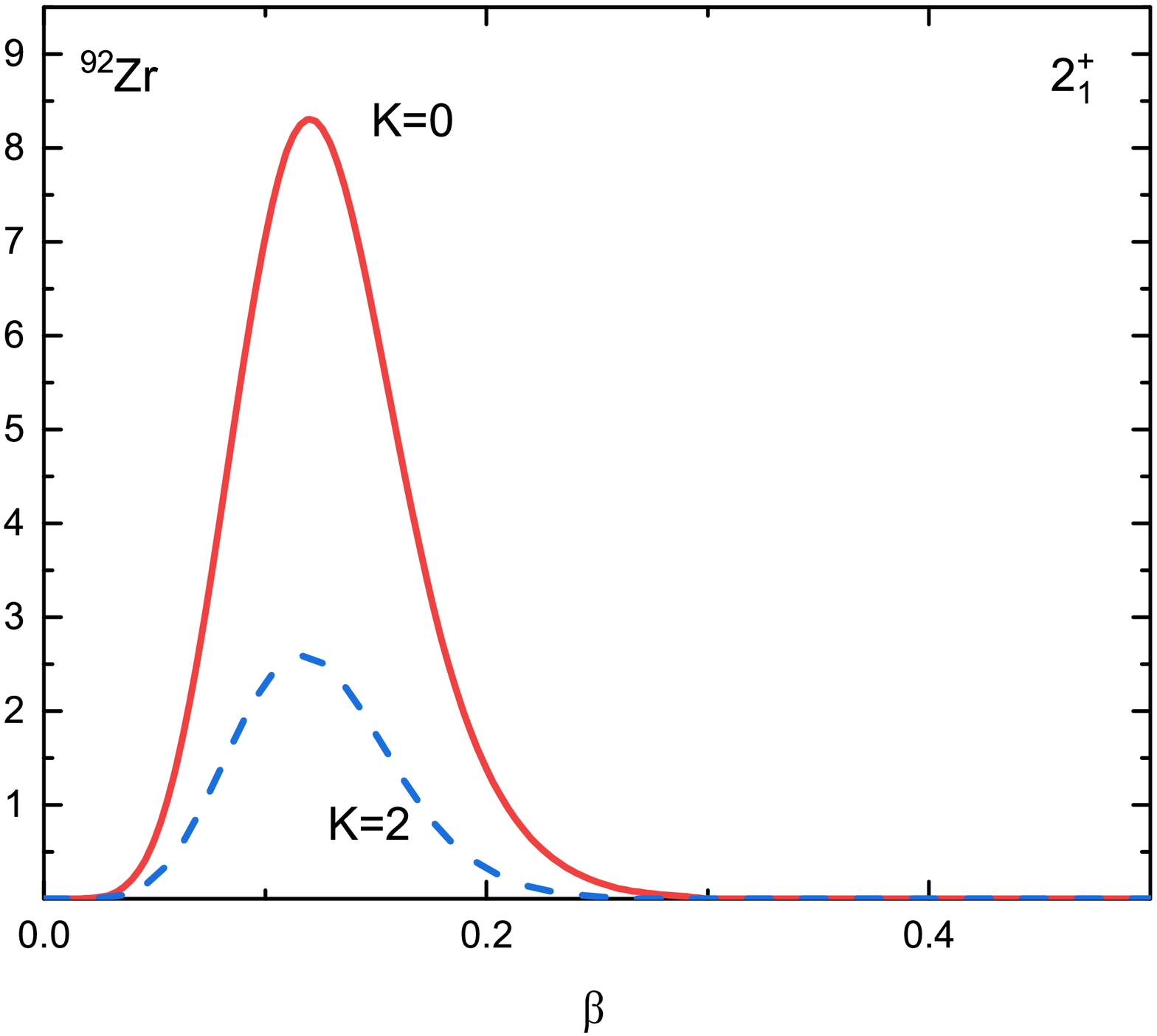}
(c)
\includegraphics
[width=0.3\textwidth]{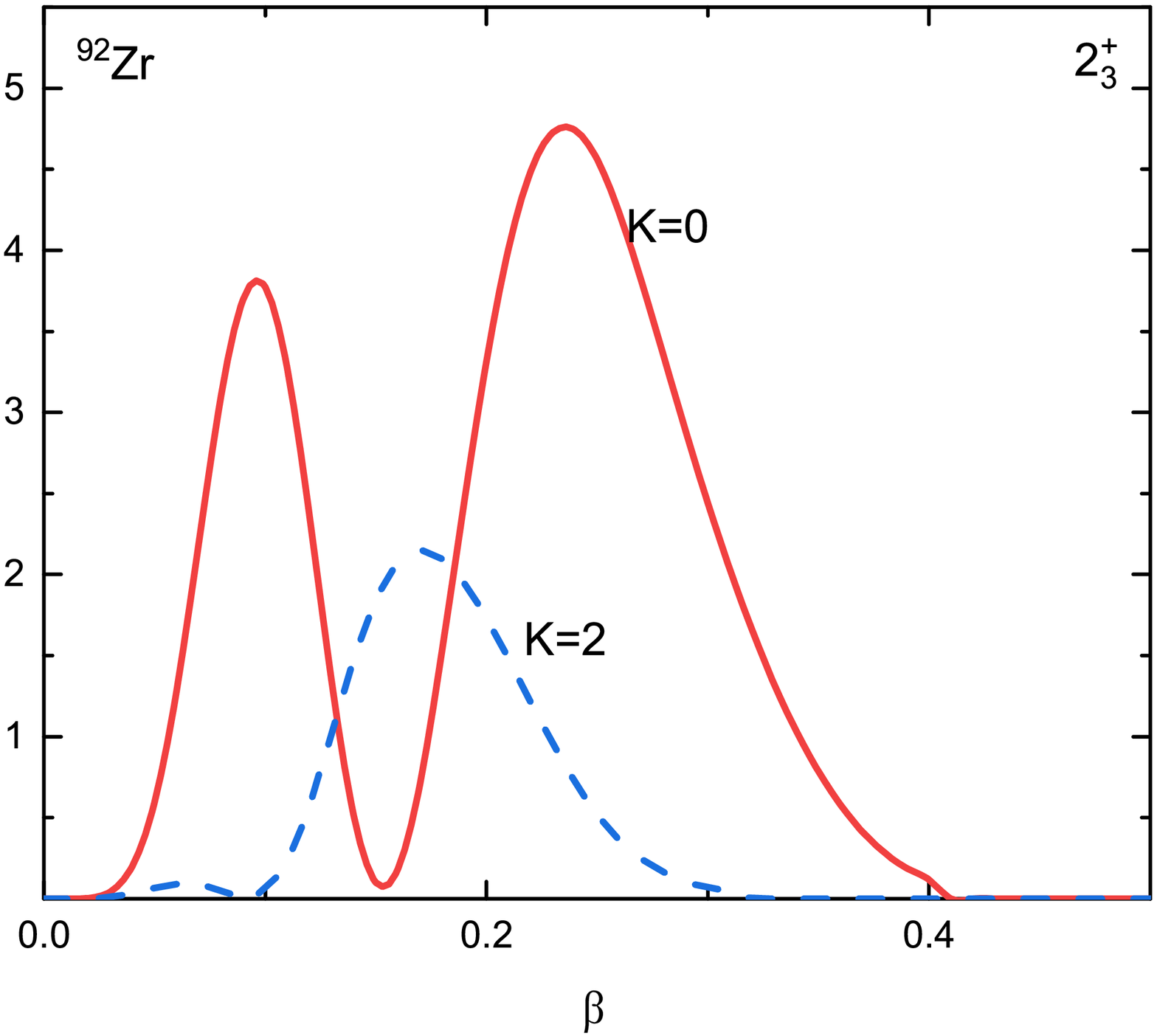}
\caption{\label{Fig1}Distribution over $\beta$ of the squares of the wave functions or their components in  $^{92}$Zr.
(a) $0^+_1$ (red short dashed line) and $0^+_2$ (purple dash-dotted line) states .
(b)$2^+_1$ state (K=0 - red solid line, K=2 - blue dashed line).
(c) $2^+_3$ state (K=0 - red solid line, K=2 - blue dashed line). }
\end{figure*}
\subsection{$^{92}$Zr}

Let's take a closer look at the  results of calculations for $^{92}$Zr. In this nucleus, the value of the $B(E2;2^+_1\rightarrow0^+_1)$ of 6.4 W.u. indicates the collective nature of the $2^+_1$ state related to the  oscillations of the nuclear shape around the spherical one.  Of course, $^{92}$Zr is not the right nucleus to consider in a collective model. It has only two valence nucleons, and even taking into account a possible transition of the pair of protons from $p_{1/2}$ to $g_{9/2}$ single partical levels, this increases the number of valence nucleons to six only. Nevertheless, we include this nucleus into consideration, since our main task is to obtain the most complete information on the evolution of the collective potential along the Zr isotope chain. In the Tables I and II  the experimental data and the results of our calculations for the low-lying states of $^{92}$Zr are compared . The calculated value of the $B(E2;2^+_1\rightarrow0^+_1)=8.2$ W.u. is close to the experimental value 6.4 W.u. Comparison of the experimental $E2$ reduced transition probabilities from $2^+_2$ (1847 keV) and $2^+_3$ (2067 keV) states to $2^+_1$ state with the calculated ones shows that the calculated state with excitation energy 1770 keV is close in properties to the experimentally observed  $2^+_3$ (2067 keV) state. That's why in the tables   the calculated state with the excitation energy of 1770 keV is compared with the observed  $2^+_3$ state.  Thus, we have the $B(E2;2^+_3\rightarrow2^+_1)_{exp}\leqslant 15$ W.u. and the $B(E2;2^+_3\rightarrow2^+_1)_{cal}=16.3$ W.u. Further, the $B(E2;2^+_3\rightarrow0^+_1)_{exp} < 0.004$ W.u and the $B(E2;2^+_3\rightarrow0^+_1)_{cal}=0.02$ W.u. This comparison indicates the collective nature of the $2^+_3$ state, which is confirmed by the small value of the $B(M1;2^+_3\rightarrow 2^+_1)$. A small value of the $B(E2;2^+_2\rightarrow2^+_1)_{exp}=0.001$ W.u. and large value of the $B(M1;2^+_2\rightarrow2^+_1)_{exp}$ indicates on the noncollective nature of the experimentally observed $2^+_2$ state. Therefore, description of the observed $2^+_2$ state requires  the shell model consideration and is not considered in this paper. For this reason the results of calculations of the $E2$ reduced transition probabilities from the $2^+_2$ state and to this state are not given in Table II.

The experimental value of the $B(E2;0^+_2\rightarrow 2^+_1)_{exp} =14.4$ W.u. indicates on the collective nature of the $0^+_2$ state. From this point of view  it is in a correspondence with the calculated value of the $B(E2;0^+_2\rightarrow 2^+_1)_{cal} =20.7$, Although the calculated value is significantly larger than the experimental one. Comparison of the experimental and calculated values of the $B(E2;4^+_1\rightarrow2^+_1)$: 4.05 W.u. and 17.7 W.u., correspondingly, indicates on the presence in the wave function of the $4^+_1$ state  of the shell model configuration $(d_{5/2}d_{5/2})_4$. This explains the strong discrepancy between the experimental and calculated results. The proximity in the value of the experimental E2 reduced  transition probabilities $B(E2;4^+_1\rightarrow 2^+_1)_{exp}$ and $B(E2;4^+_2\rightarrow 2^+_1)_{exp}$ indicates that the lowest energy pure collective $4^+$ state generated by the Bohr collective Hamiltonian  is fragmented, and distributed between the experimentally observed $4^+_1$ and $4^+_2$ states. This statement is consistent with the fact that the calculated value of $B(E2;4^+_2\rightarrow 2^+_1)$ is practically zero.

In Fig.~\ref{Fig1}  a $\beta$-dependence of the wave functions of the $2^+_1$, $0^+_2$ and $2^+_3$ states of $^{92}$Zr is demonstrated. It is seen that the wave functions of the $0^+_2$ and $2^+_3$ states have a similar $\beta$-dependance, namely, they have a node in $\beta$ in contrast to the $2^+_1$ state. This makes these states similar to the two-phonon states, which explains the large value
 of the  $E2$ reduced transition probability from these states to the $2^+_1$ state. Our calculations qualitatively confirm these experimental date.

\begin{figure*}[hbt]
\centering
(a)
\includegraphics
[width=0.3\textwidth]{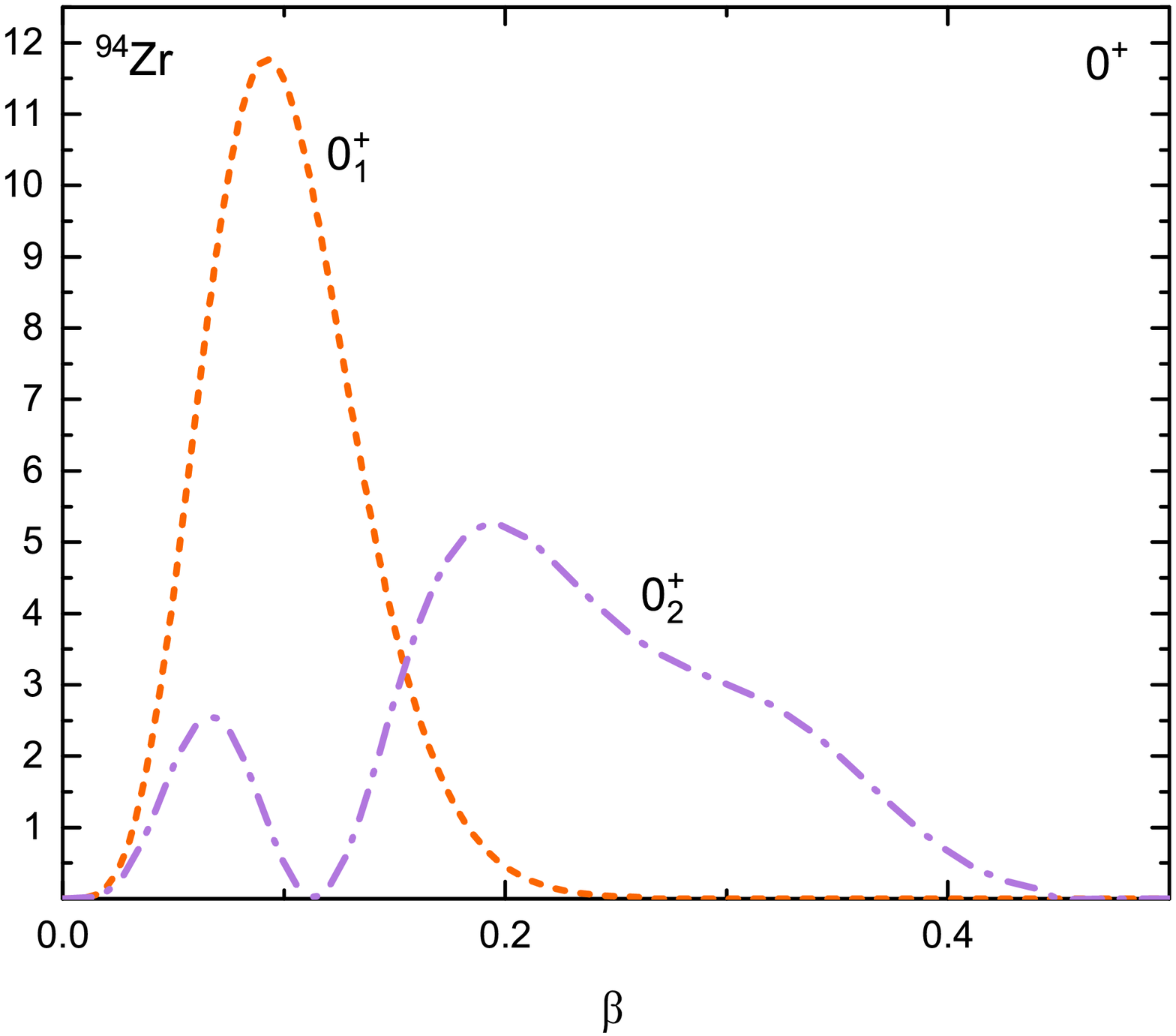}
(b)
\includegraphics
[width=0.3\textwidth]{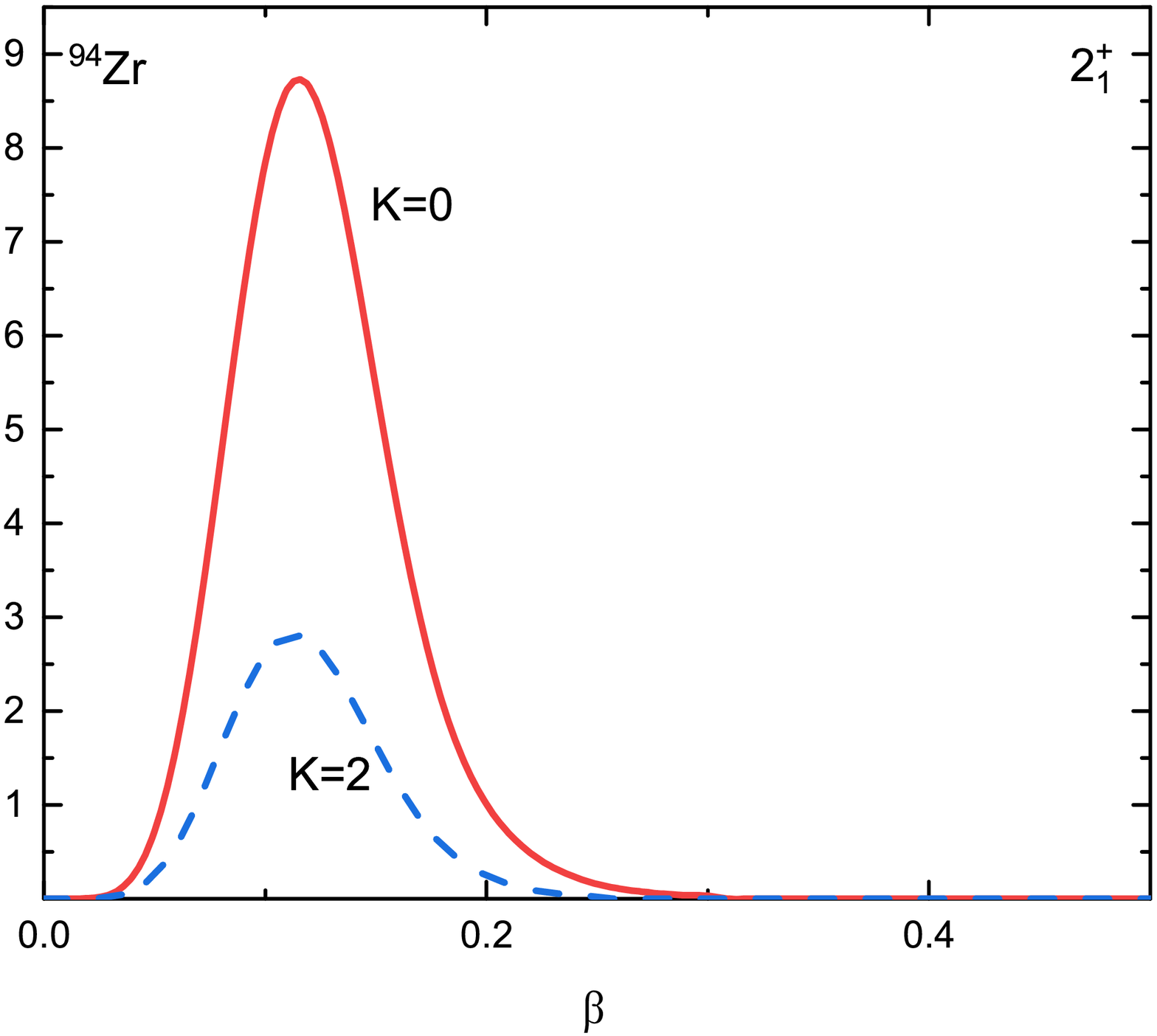}
(c)
\includegraphics
[width=0.3\textwidth]{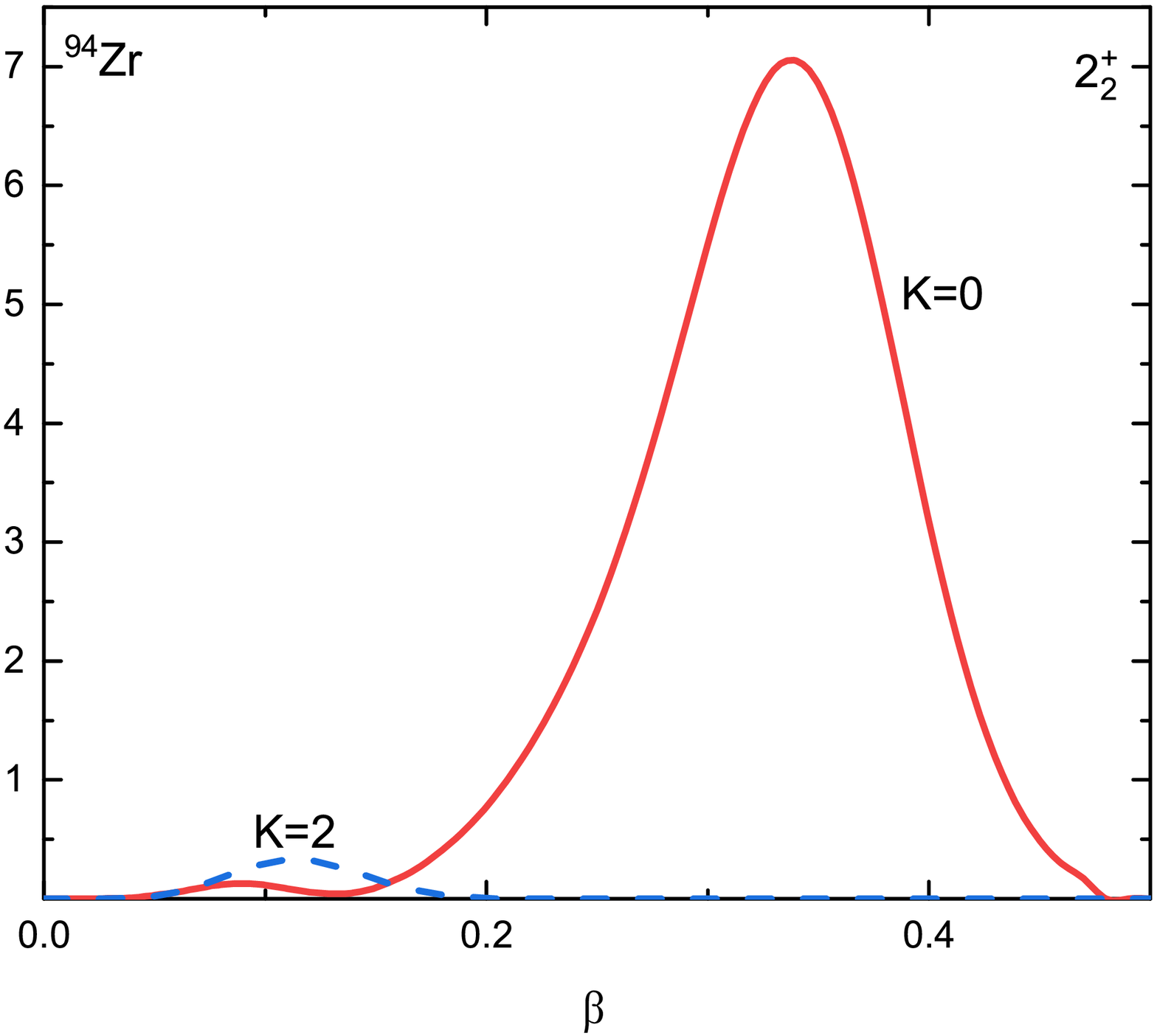}
(d)
\includegraphics
[width=0.3\textwidth]{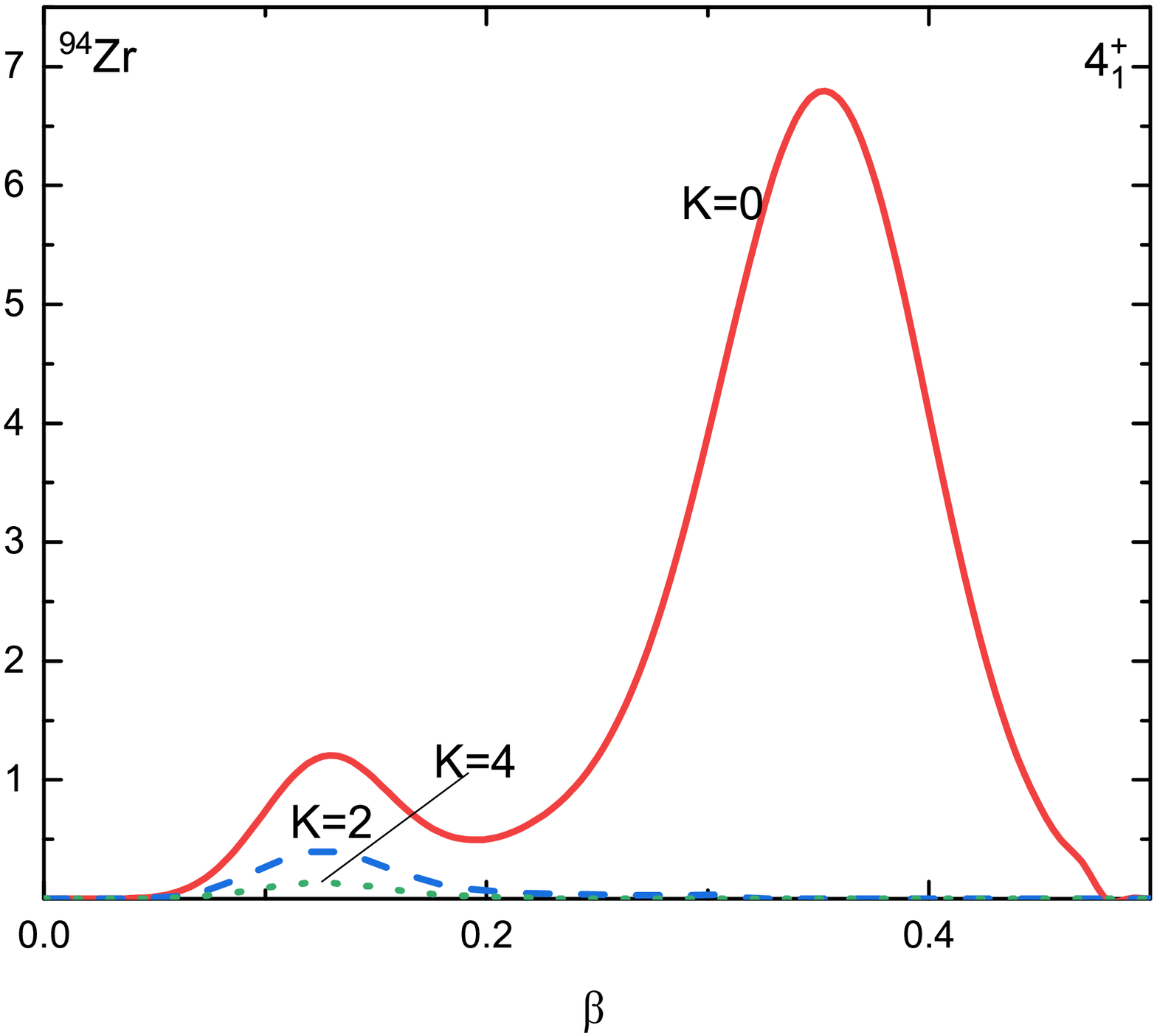}
(e)
\includegraphics
[width=0.3\textwidth]{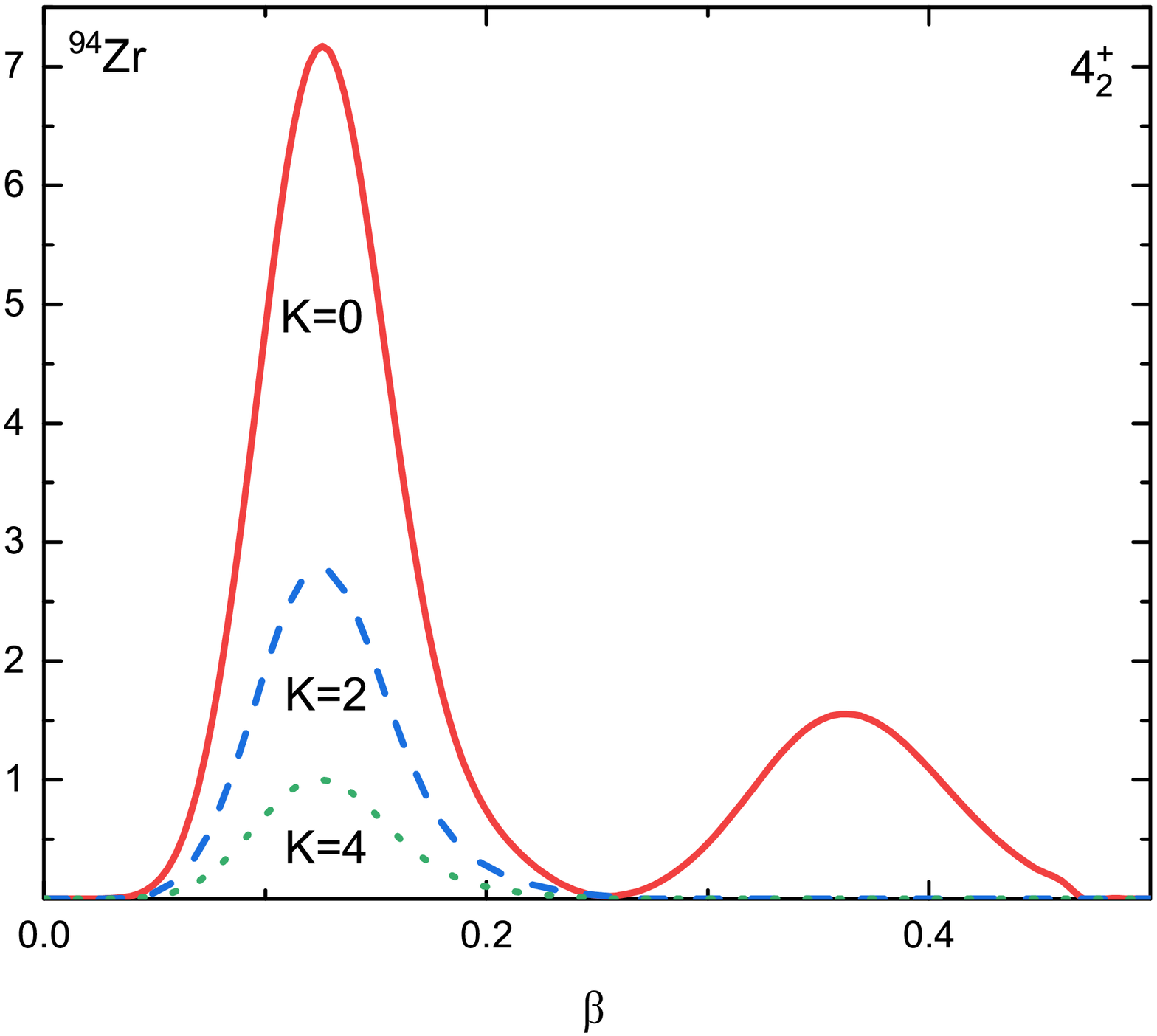}
\caption{\label{Fig2f5}
Distribution over $\beta$ of the squares of the wave functions or their components in $^{94}$Zr.
(a)  $0^+_1$ (red short dashed line) and $0^+_2$ (purple dash-dotted line) states.
(b) $2^+_1$ state (K=0 - red solid line, K=2 - blue dashed line).
(c) $2^+_2$ state (K=0 - red solid line, K=2 - blue dashed line).
(d) $4^+_1$ state (K=0 - red solid line, K=2 - blue dashed line, K=4 - blue).
(e) $4^+_2$ state (K=0 - red solid line, K=2 - blue dashed line, K=4 - green dotted line). }
\end{figure*}
\subsection{$^{94}$Zr}

Let us consider the experimental data on $^{94}$Zr. The experimental value of  $B(E2;2^+_1\rightarrow0^+_1)=4.9$ W.u. indicates that the  $2^+_1$ state of this nucleus is quite collective, and its structure is formed by excitation of both proton and neutron subsystems. However, the predominance of the neutron shell model configuration $(d_{5/2}d_{5/2})_2$, whose contribution to the value of the $B(E2;2^+_1\rightarrow0^+_1)$ is small, cannot be excluded. This possibility is indicated by the value of $B(E2;2^+_1\rightarrow0^+_1)_{cal}=8.09$ W.u. calculated using the standard definition of the quadrupole moment operator in the Geometrical Collective Model. Comparing this value with the experimental one, it can be assumed that the contribution of a noncollective component to the wave function of the $2^+_1$ state may be of the order of 50\%.

The experimental value of  $B(E2;4^+_1\rightarrow 2^+_1)_{exp}$ is noticeably less than $B(E2;2^+_1\rightarrow 0^+_1)_{exp}$, which indicates or on a predominance of the noncollective component $(d_{5/2}d_{5/2})_4$ in the structure of the $4^+_1$ state, either on a deformed nature of the $4^+_1$ state in contrast to the spherical $2^+_1$ state. Our calculations support a second possibility. At the same time a large experimental value of the $B(E2;4^+_2\rightarrow 2^+_1)_{exp}=13.2$ W.u. indicates that $4^+_2$ state  is the spherical one as  the $2^+_1$ state. This interpretation is confirmed by the $\beta$-dependence of the wave function of the $4^+_2$ state (see Fig.\ref{Fig2f5}e) which is concentrated at low $\beta$, as the wave function of the $2^+_1$ state  and has a node in $\beta$. The wave function of the $4^+_1$ state is concentrated at large $\beta$ (see Fig.\ref{Fig2f5}d). This explains the small value of the $B(E2;4^+_1 \rightarrow2^+_1)$. Our calculations of the $B(E2;4^+_1\rightarrow 2^+_1)$  confirm the experimentally observed tendency to decrease $B(E2;4^+_1\rightarrow2^+_1)$ compare to $B(E2;2^+_1\rightarrow 0^+_1)$  although give a value exceeding the experimental one.

Interpretation of the properties of the $2^+_2$ state is complicated. The large experimental value of $B(E2;4^+_2\rightarrow 2^+_2)$ qualitatively reproduced by our calculations (however, twice as less) assumes a spherical nature of the $2^+_2$ state in agreement with a spherical nature of the $4^+_2$ state. At the same time, a small experimental value of the $B(E2;2^+_2\rightarrow 2^+_1)$ contradicts to the assumption of the spherical structure of the $2^+_2$ state, because $2^+_1$ state is spherical as it is seen in Fig.\ref{Fig2f5}b. As it is shown in Fig. \ref{Fig2f5}c the $2^+_2$ state is deformed and the relatively large value of $B(E2;4^+_2\rightarrow 2^+_2)$ is explained by the presence of the second maximum of the wave function of the $4^+_2$ state at large values of $\beta$. Deformed structure of the
$2^+_2$ state  explains a small value of the $B(E2;2^+_2\rightarrow 2^+_1)$. The calculated value of $B(E2;2^+_2\rightarrow 0^+_1)$ is small in agreement with a deformed structure of the $2^+_2$ state and spherical nature of the $0^+_1$ state. However, the fact that the calculated value of
 the $B(E2;2^+_2\rightarrow 0^+_1)$ is significantly less than the experimental value indicates that the calculated $2^+_2$ state  too much shifted into deformed region.

 A large value of $B(E2;0^+_2\rightarrow 2^+_1)$ indicates that $0^+_2$ state can be similar in its structure to the two-phonon state, although modified  by the anharmonic effects. At the same time, a  large value of the $B(E2;2^+_2\rightarrow 0^+_2)$ which is also reproduced by our calculations, is rather consistent with the deformed structure of the $0^+_2$ state. In Fig.\ref{Fig2f5}  the square of the wave function of the $0^+_2$ state is presented as a function of $\beta$. It is seen, that from one side the wave function of the $0^+_2$ state has a node as a function of $\beta$, which makes the $0^+_2$ state similar to the two-phonon state. This explain a large value of the $B(E2;0^+_2\rightarrow 2^+_1)$. But from the other side, a large part of the wave function of the $0^+_2$ state is located in the region of large $\beta$. It explains the large value of the $B(E2;2^+_2\rightarrow 0^+_2)$.

\subsection{$^{96}$Zr}

The nucleus $^{96}$Zr was considered in details in our previous publication \cite{Mardyban}, albeit with slightly different parametrisation of the potential energy. The results shown in Tables I and II are close to those shown in \cite{Mardyban}. Some of the values calculated in this work are closer to the experimental values than those obtained in \cite{Mardyban} . In several cases, previous results are better agreed with experimental data. This applies to  $B(E2;4^+_1\rightarrow 2^+_1)$ and  $B(E2;4^+_2\rightarrow2^+_2)$.

\begin{figure*}[htb]
\centering
(a)
\includegraphics
[width=0.3\textwidth]{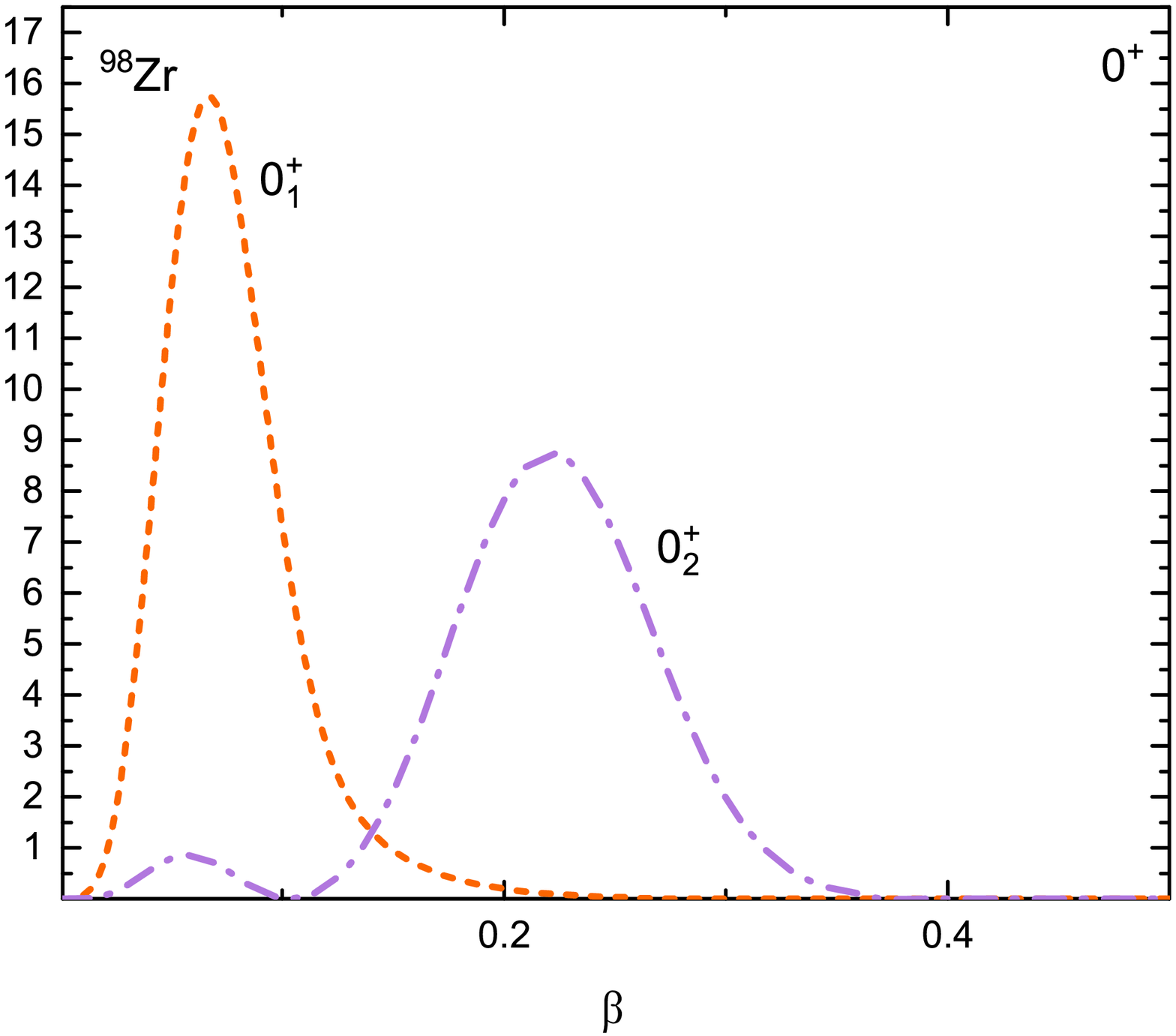}
(b)
\includegraphics
[width=0.3\textwidth]{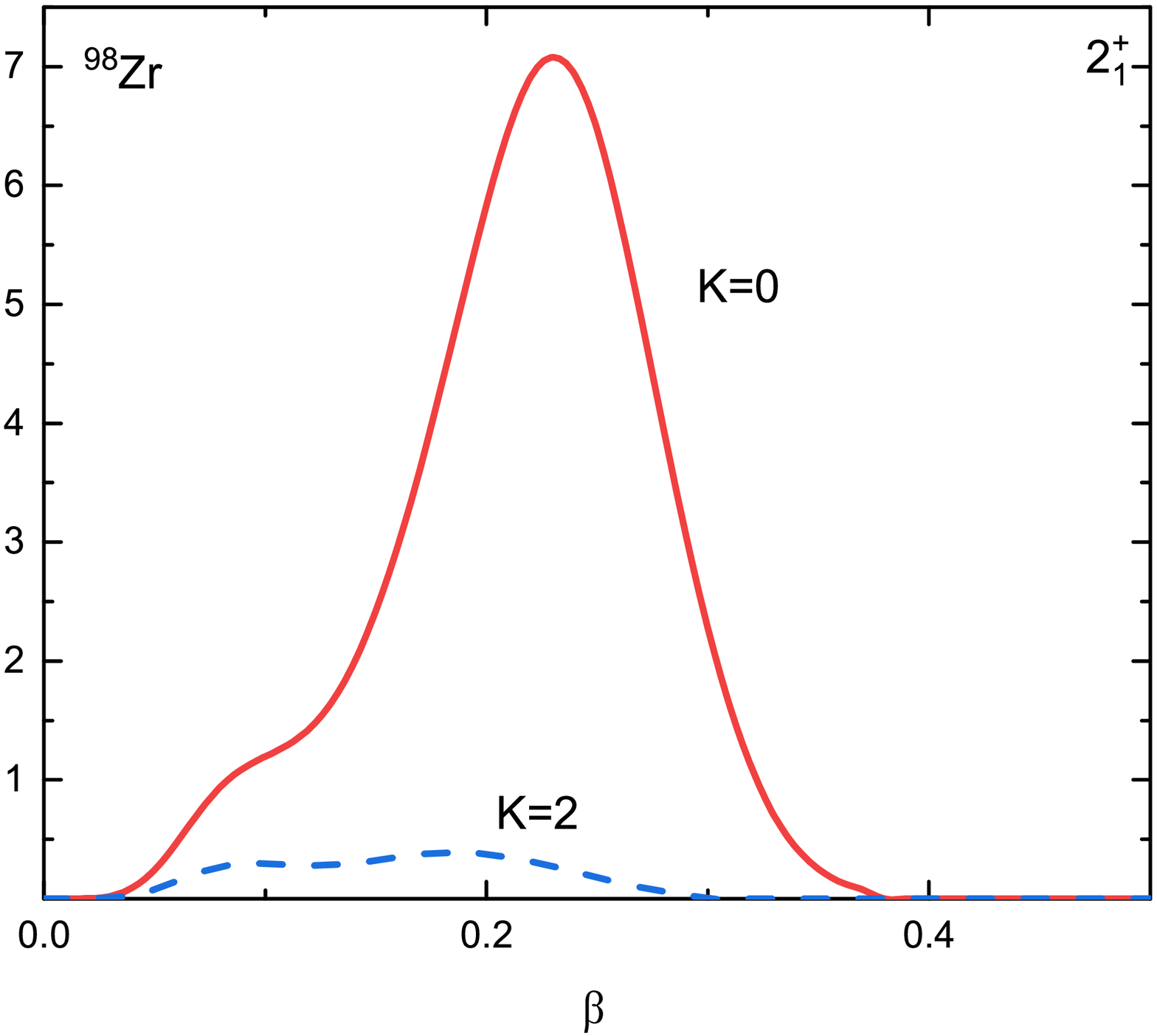}
(c)
\includegraphics
[width=0.3\textwidth]{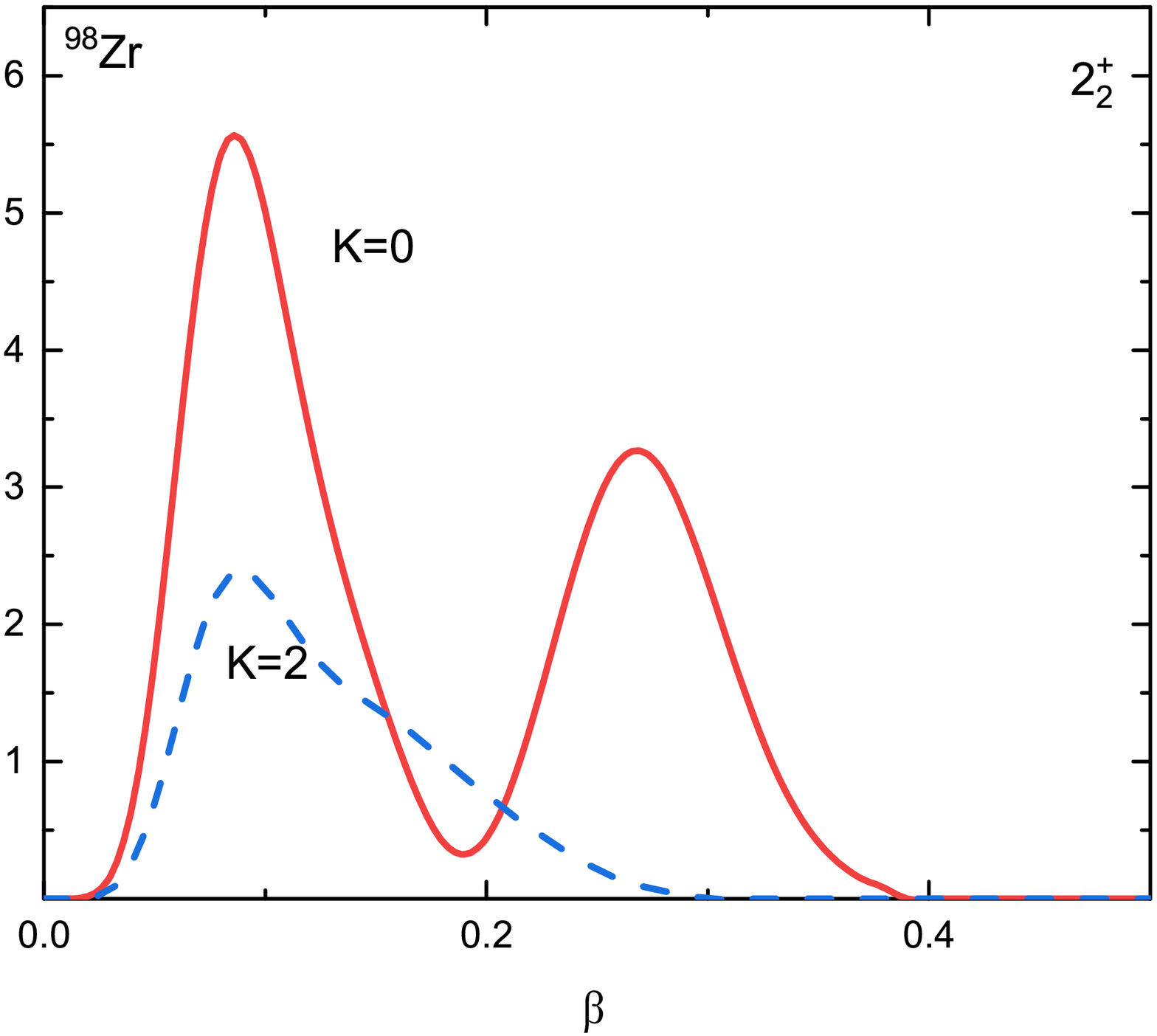}
(d)
\includegraphics
[width=0.3\textwidth]{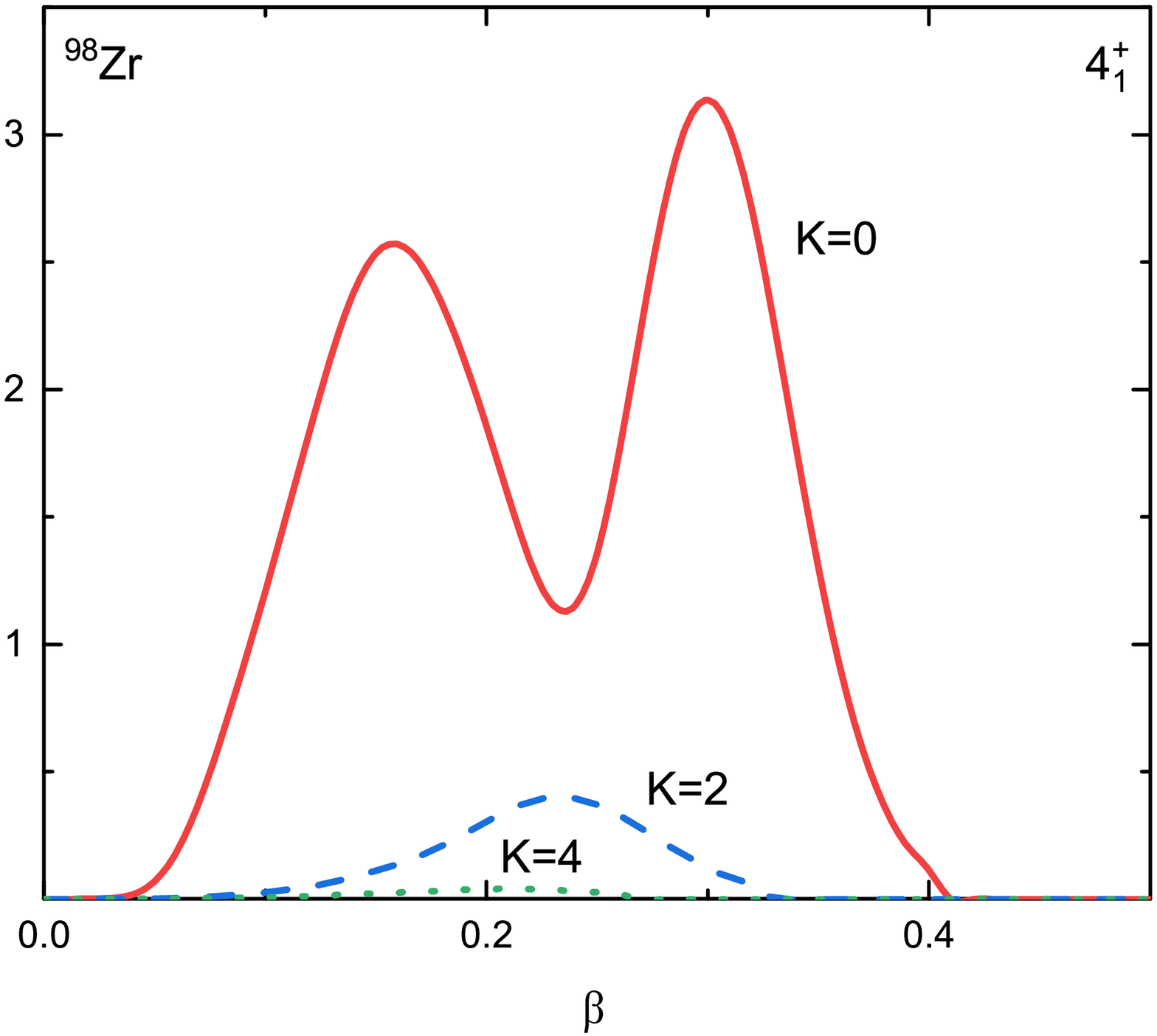}
(e)
\includegraphics
[width=0.3\textwidth]{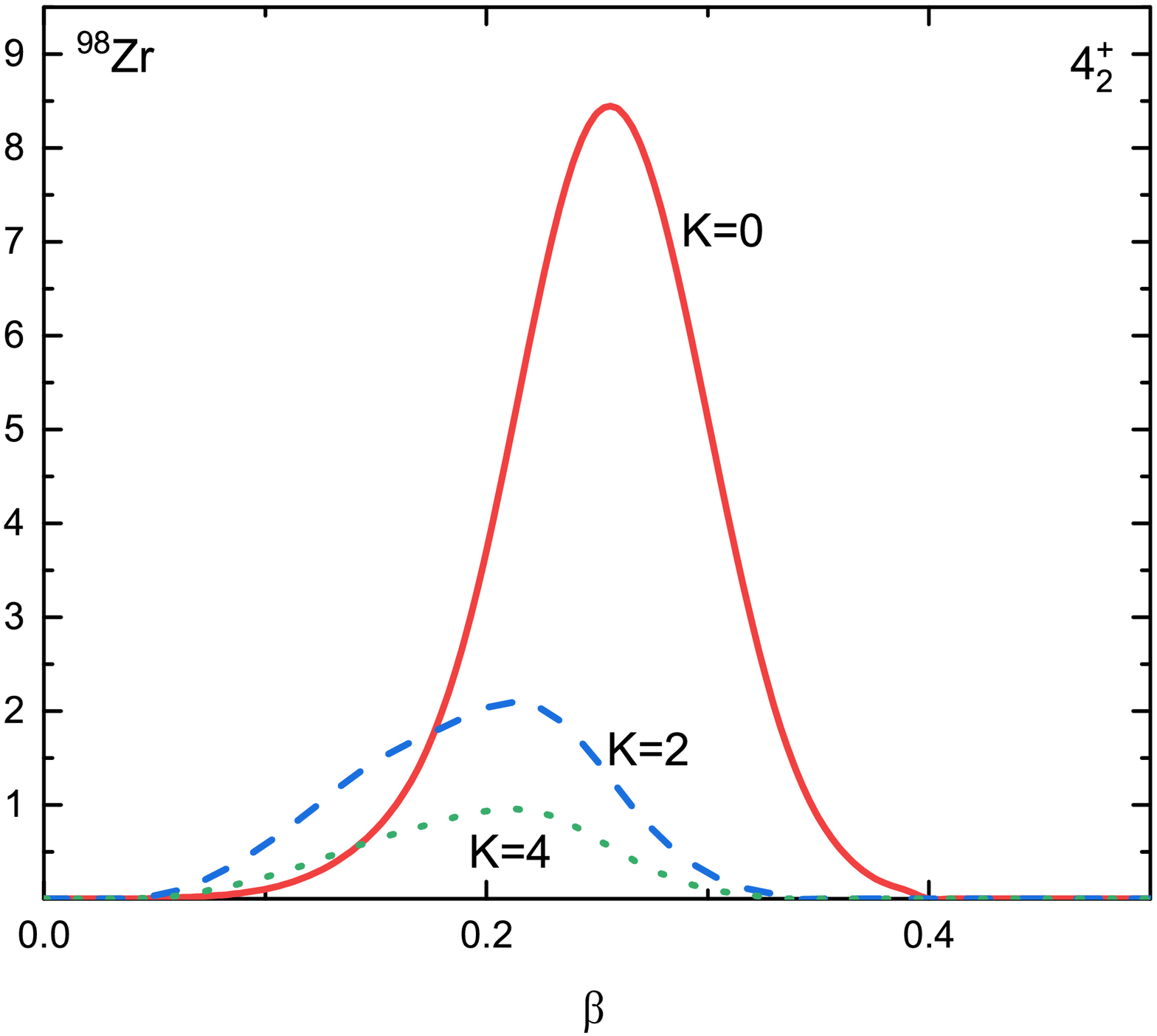}
\caption{\label{Fig3}
The same as in Fig.\ref{Fig2f5}, but for $^{98}$Zr.}
\end{figure*}
\subsection{$^{98}$Zr}

The wave functions of the low lying states of $^{98}$Zr are shown in Fig.\ref{Fig3}. It is seen that the wave function of the $0^+_1$ state is concentrated around
$\beta\approx 0.1$, i.e. in the spherical region, but the $0^+_2$ state is deformed.

The wave function of the $2^+_1$ state has a wide maximum centered at $\beta=0.27$, i.e. is distributed over both spherical and deformed regions. A similar structure has a wave function of the $4^+_2$ state. The wave functions of the $2^+_2$  and $4^+_1$ states are similar to each other. The squares of their wave functions has two maxima: one spherical and the other deformed.

The following $E2$ transitions are measured:  $B(E2;0^+_2\rightarrow 2^+_1)_{exp}=29.0$ W.u. and  $B(E2;2^+_1\rightarrow 0^+_1)_{exp}=2.9$ W.u.
The large value of the $B(E2;0^+_2\rightarrow 2^+_1)_{exp}$ is explained by the fact that the wave functions of both $0^+_2$ and $2^+_1$ states
have a similar structure and are located in deformed region. The relatively small value of the $B(E2;2^+_1\rightarrow 0^+_1)_{exp}$ is explained by the fact
that the wave function of the $0^+_1$ state is concentraited in a spherical region, in contrast to the $0^+_2$ state.

A large experimental value of the $B(E2;4^+_1\rightarrow 2^+_1)_{exp}=42$ W.u. is reproduced in calculations: $B(E2;4^+_1\rightarrow 2^+_1)_{cal}=47$ W.u.
It is explained by the fact that the wave function of the $2^+_1$ state is located in deformed region, and the wave function of the $4^+_1$ has
a strong maximum in the same region. The calculated value of the $B(E2;4^+_1\rightarrow 2^+_2)_{cal}=11.3$ W.u. is characteristic for the collective $E2$ transition.
However, its value is five times smaller than the experimental one. Probably, this means that the maximum of the wave function of the $2^+_2$ state
located at $\beta=0.1$ should be lower than that obtained in our calculations and shown in Fig.\ref{Fig3}c. However, deformed maximum of this wave function should be higher than that shown in Fig.\ref{Fig3}c.

\subsection{$^{100,102}$Zr}

\begin{figure*}[hbt]
\centering
(a)
\includegraphics
[width=0.45\textwidth]{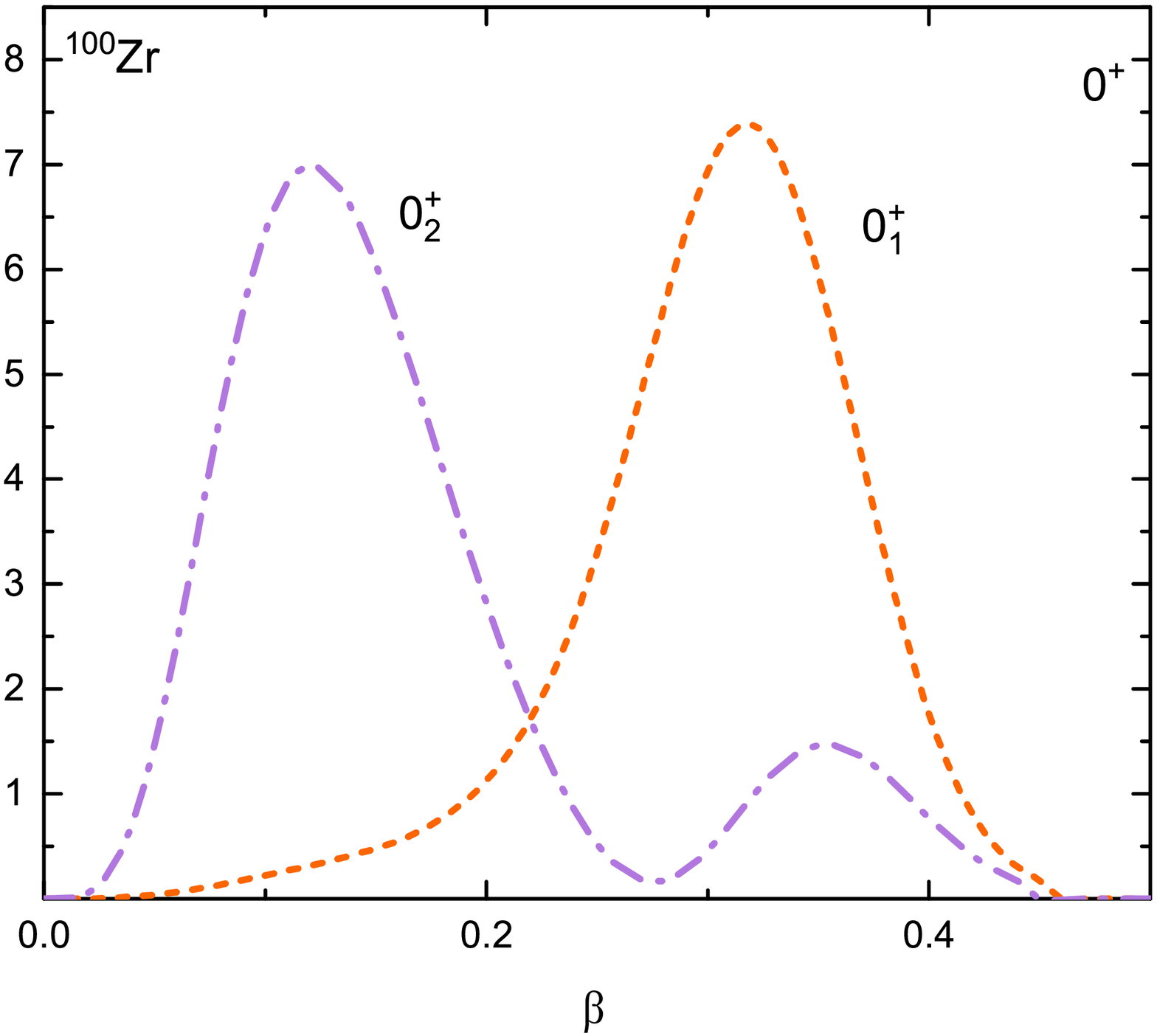}
(b)
\includegraphics
[width=0.45\textwidth]{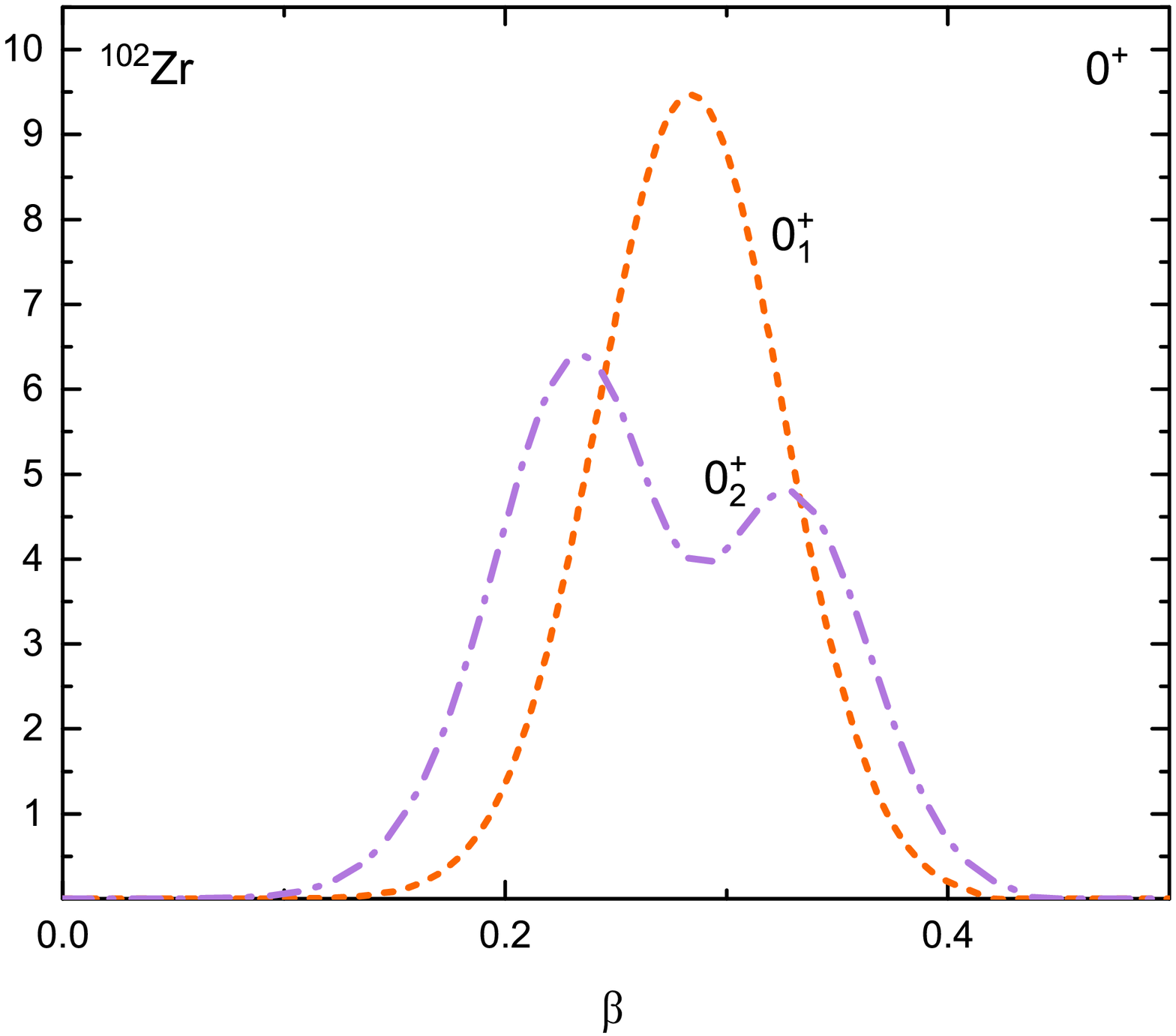}
\caption{\label{Fig4}
Distribution over $\beta$ of the squares of the wave functions of the $0^+_1$ (red short dashed line) and $0^+_2$ (purple dash-dotted line) states.
(a)  $^{100}$Zr.
(b)  $^{102}$Zr.}
\end{figure*}

In $^{100}$Zr we consider only two quasi-rotational bands based on $0^+_1$ and $0^+_2$ states. The more strong $E2$ transitions connect the states of the ground band based on the $0^+_1$. The results of calculations of these transitions are in agreement with the experimental data. Unfortunately, there are no experimental data for $E2$ transitions between the states of the second band.
The calculated values of $B(E2)$ for this band are noticeably smaller than those for transitions within the ground band. However, they are strong enough to confirm their collective nature. Calculated values of the $E2$ transitions between rotational bands are small compared to probabilities of the $E2$ transitions inside rotational bands. Based on the values of $B(E2)$ for transitions inside the bands, we conclude that the rotational band based on $0^+_1$ state is rather deformed, and that based on the $0^+_2$ is more spherical. This conclusion is supported by the calculated $\beta$-dependence of the wave functions of the $0^+_1$ and $0^+_2$ sates shown in Fig.\ref{Fig4}a.

The experimental data and the results of calculation for $^{102}$Zr present a picture similar to described above for $^{100}$Zr. In $^{102}$Zr there is a rotational band based on the ground state, with strong $E2$ transitions between the states of the band. This indicates that the $0^+_1$ state is deformed. The excited band is based on the $0^+_2$ state with the calculated excitation energy 900 keV. This band is characterized by fairly collective, but still weaker than in the case of the ground band, $E2$ transitions within the band. The calculated excitation energy of $0^+_2$ state indicates that excited band can be interpreted as a quasi-beta band. However, smaller calculated values of the $E2$ transition probabilities inside this band indicate that the $0^+_2$ wave function has an admixture of the spherical component. Wave functions of the $0^+_1$ and $0^+_2$ states are shown in Fig.\ref{Fig4}b. From Fig.\ref{Fig4}b it is seen that the wave function of the $0^+_2$ state oscillates like the wave function of the $\beta$-vibrational state. However, it is shifted to the spherical region with a greater extent than the wave function of the $0^+_1$ state.

\begin{figure*}[htb]
\begin{tabular}{llllll}
\includegraphics[scale=0.4]{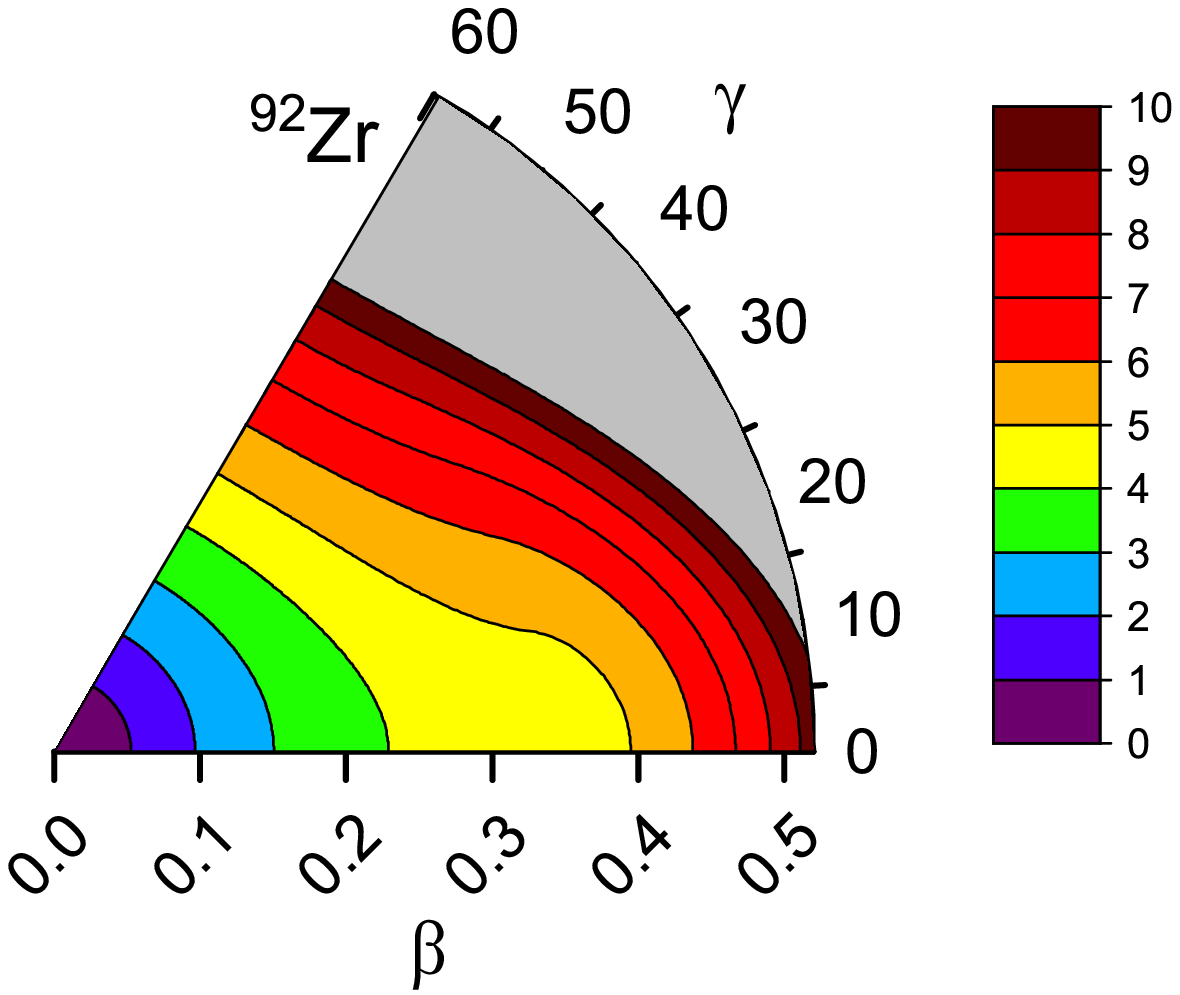}&
\includegraphics[scale=0.4]{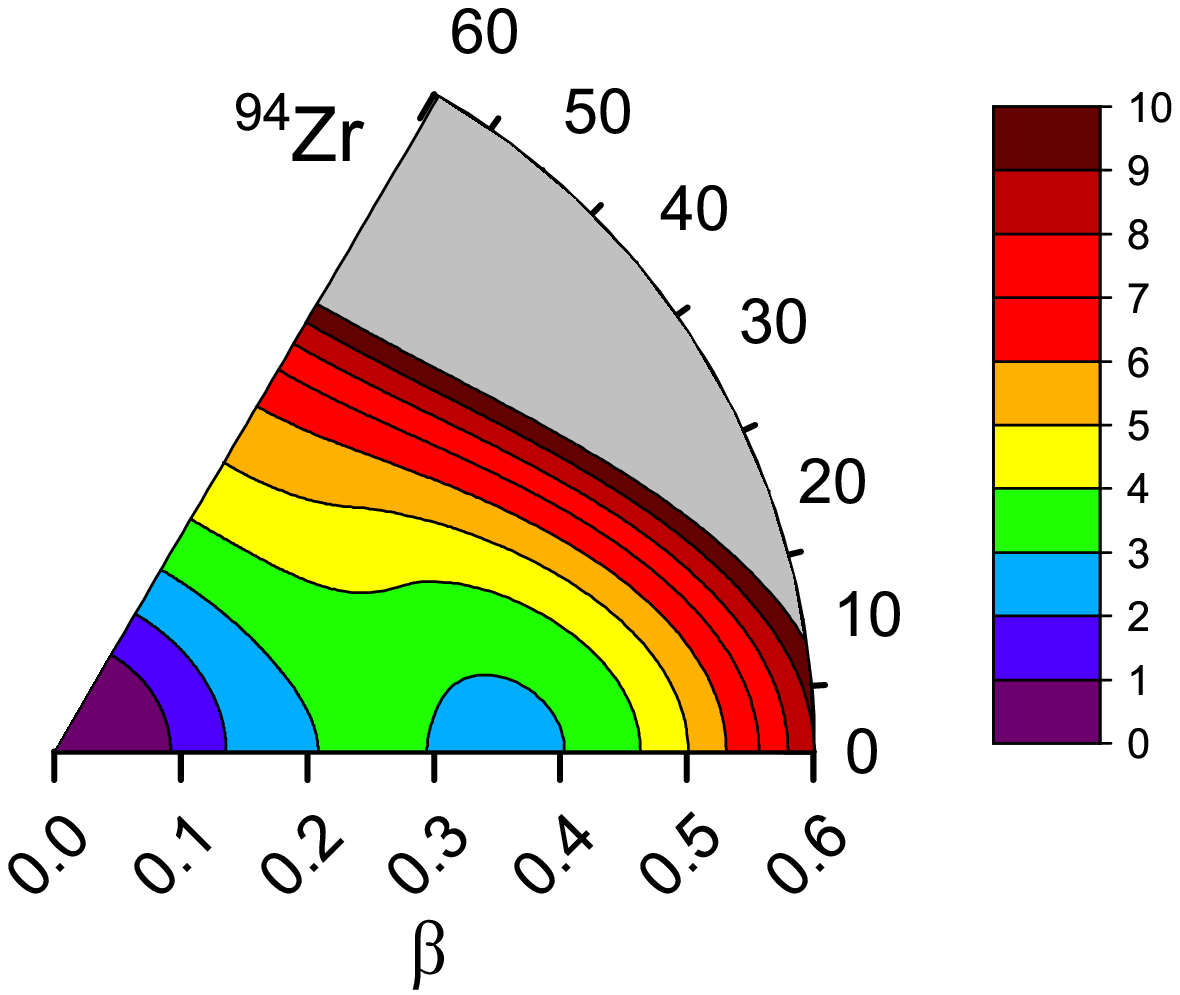}&
\includegraphics[scale=0.4]{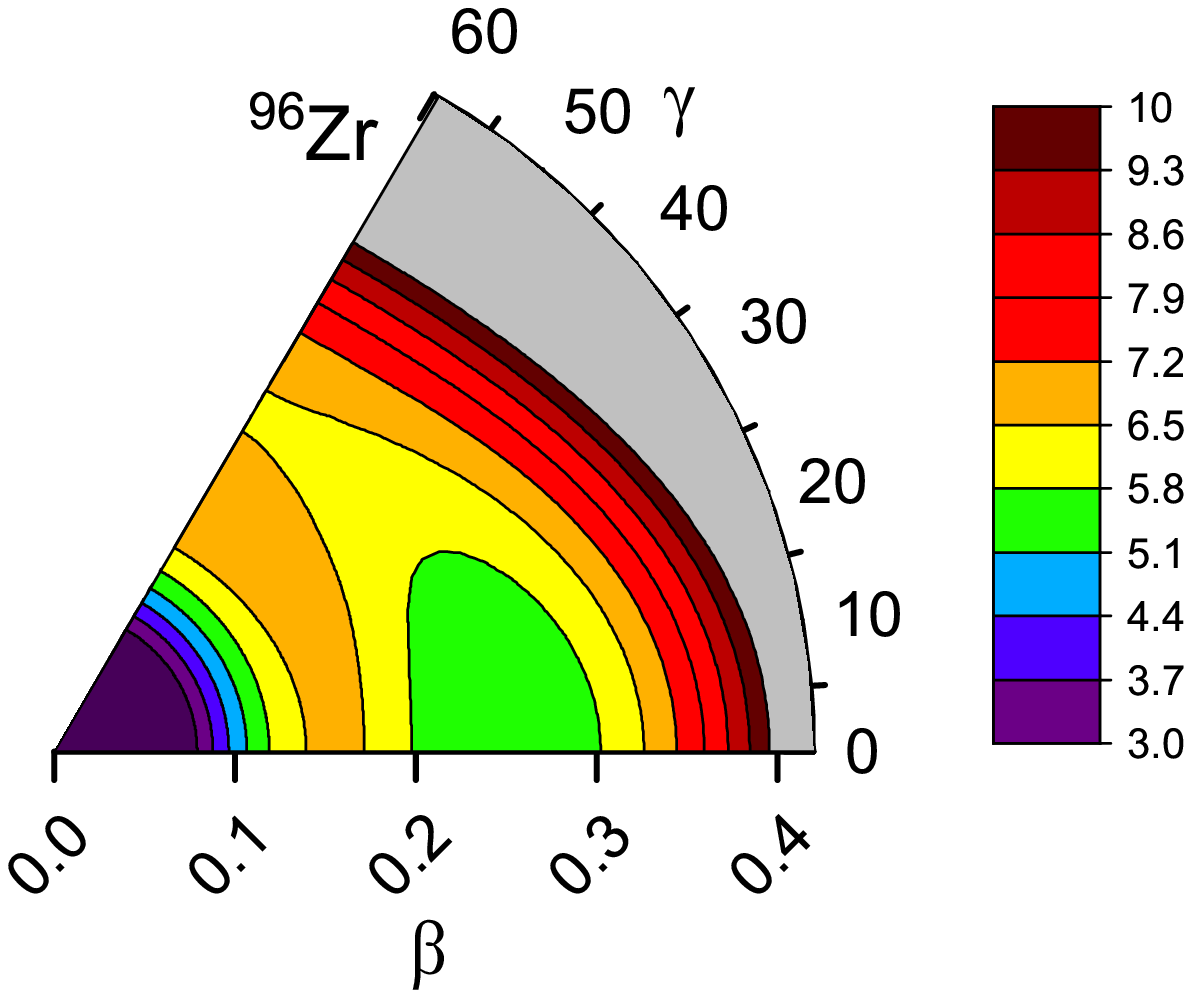} \\
\includegraphics[scale=0.2]{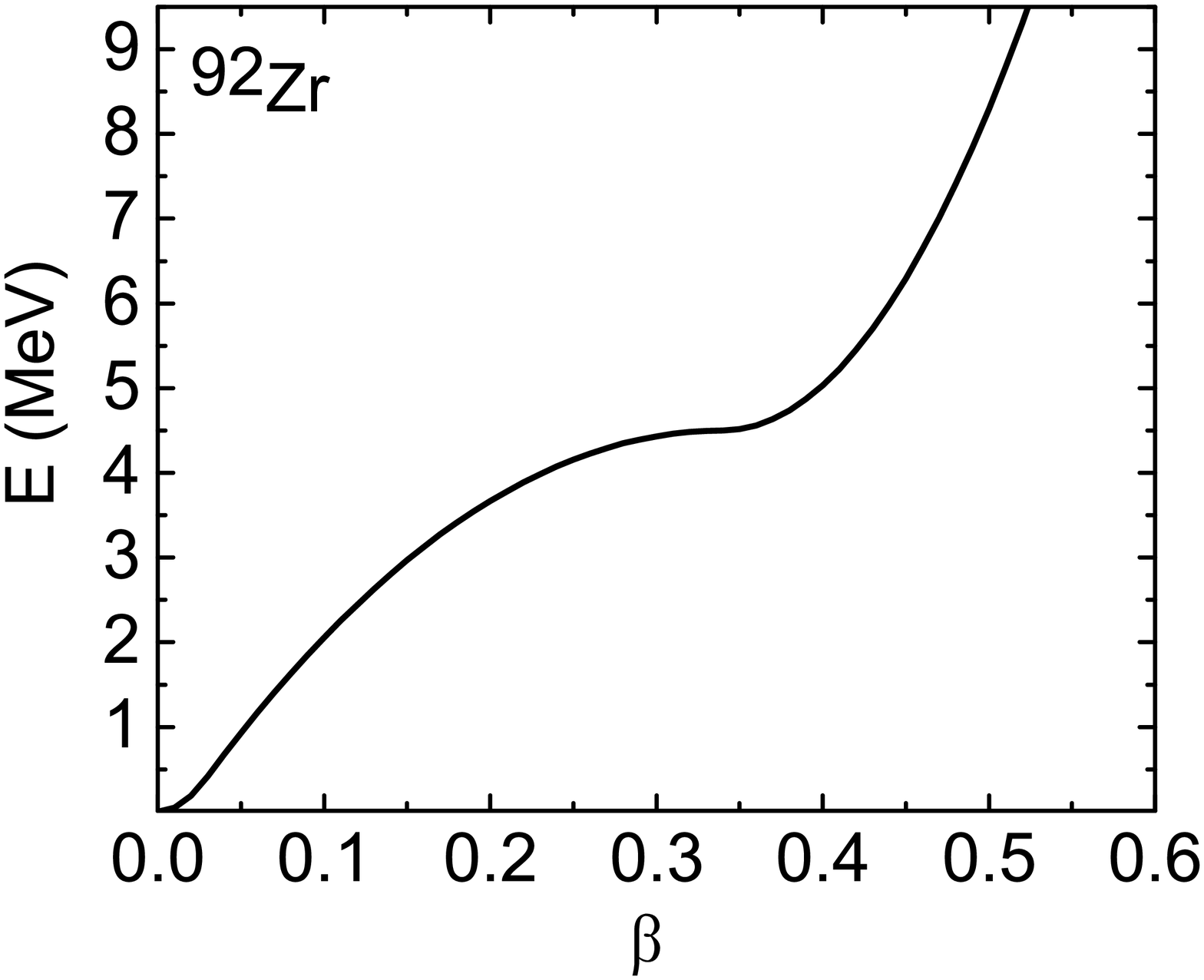}&
\includegraphics[scale=0.2]{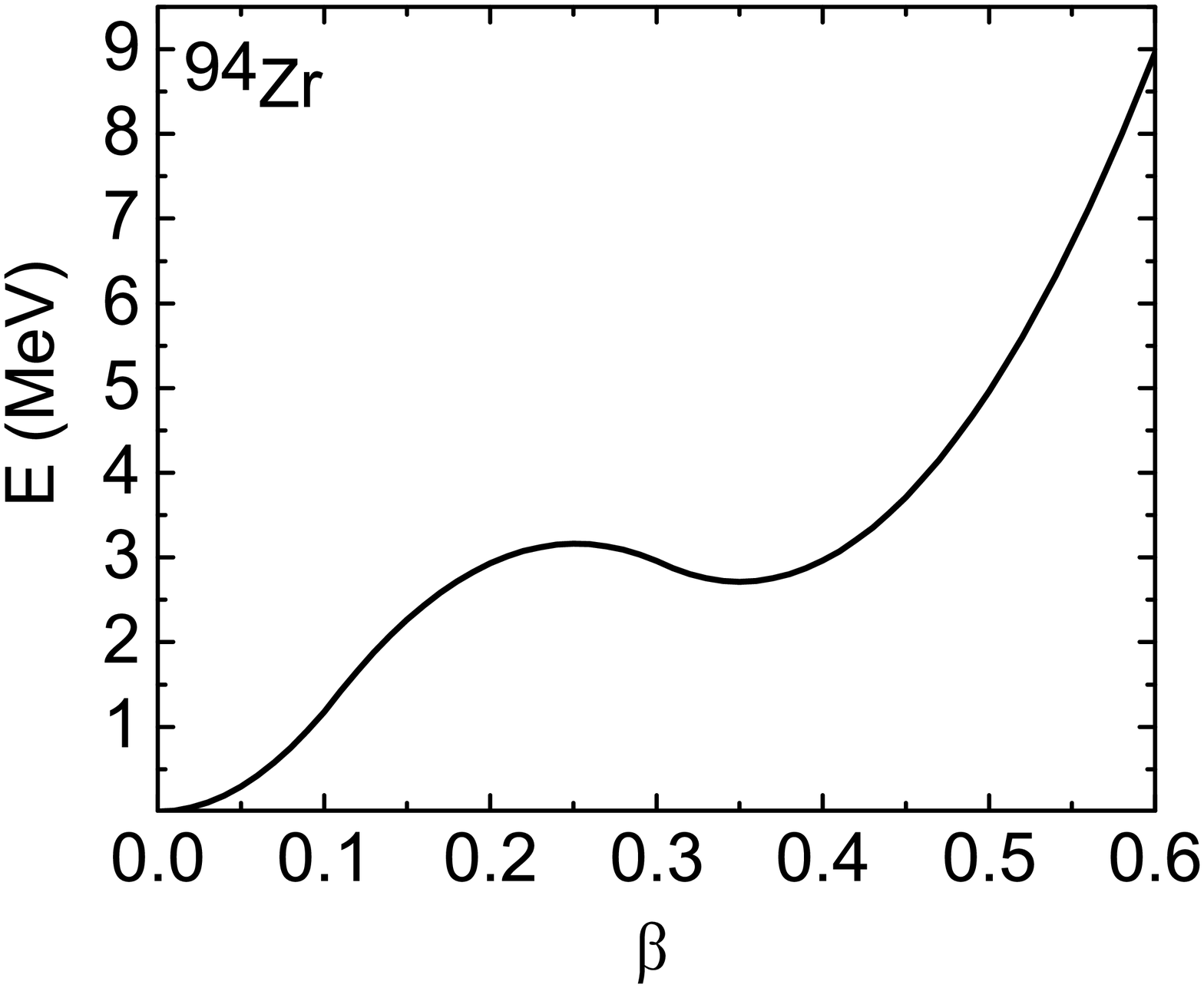}&
\includegraphics[scale=0.2]{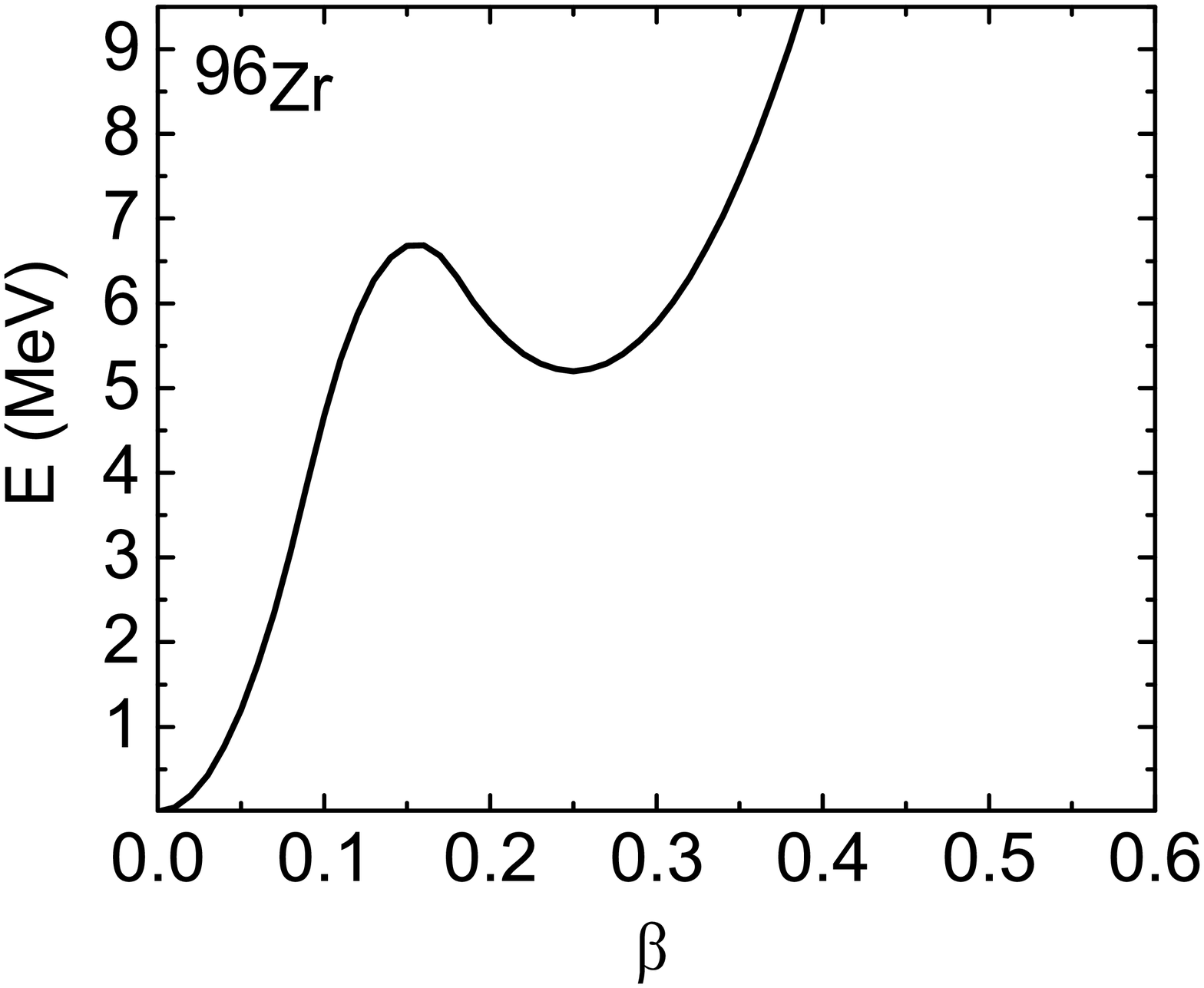} \\
\includegraphics[scale=0.4]{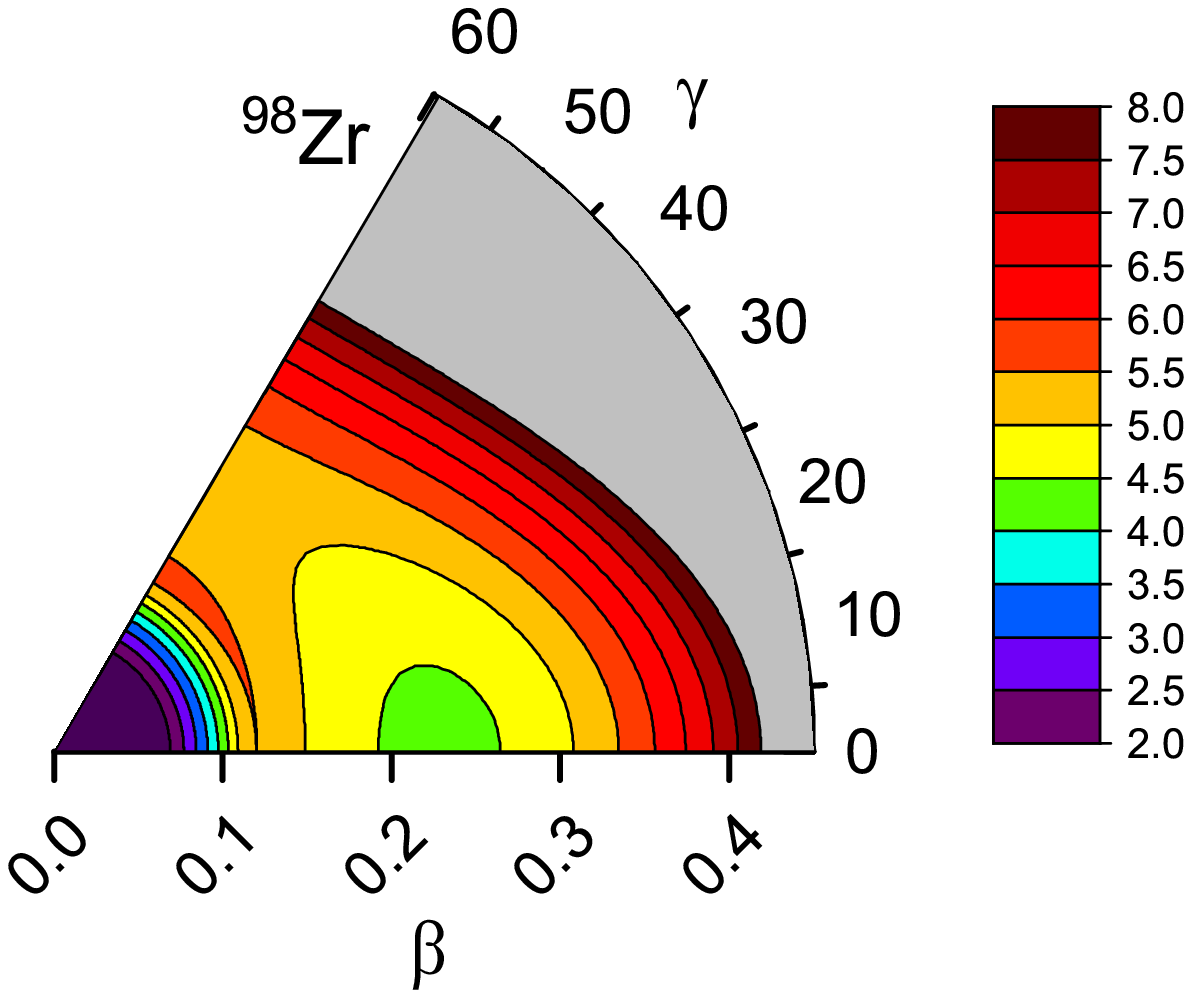}&
\includegraphics[scale=0.4]{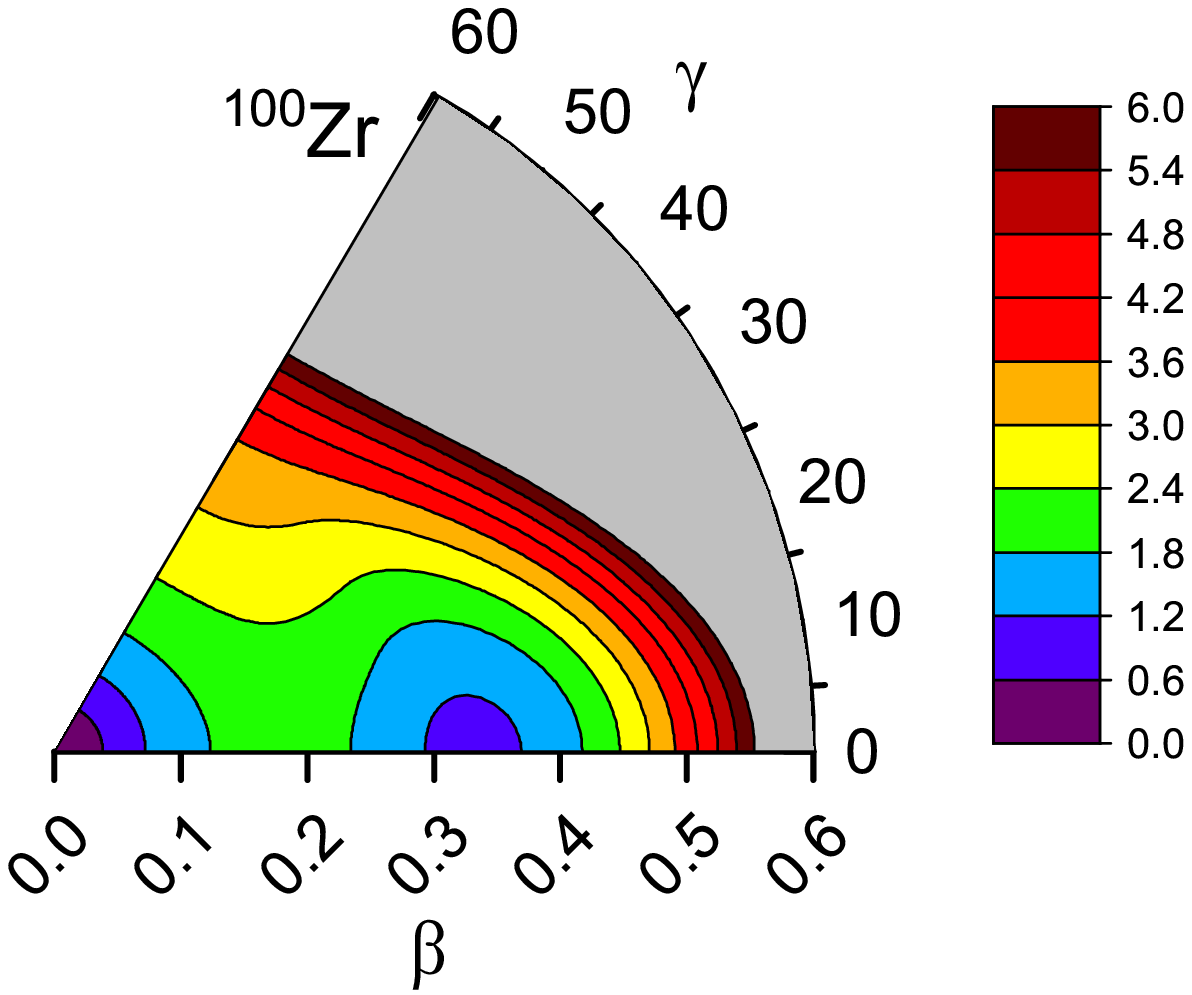}&
\includegraphics[scale=0.4]{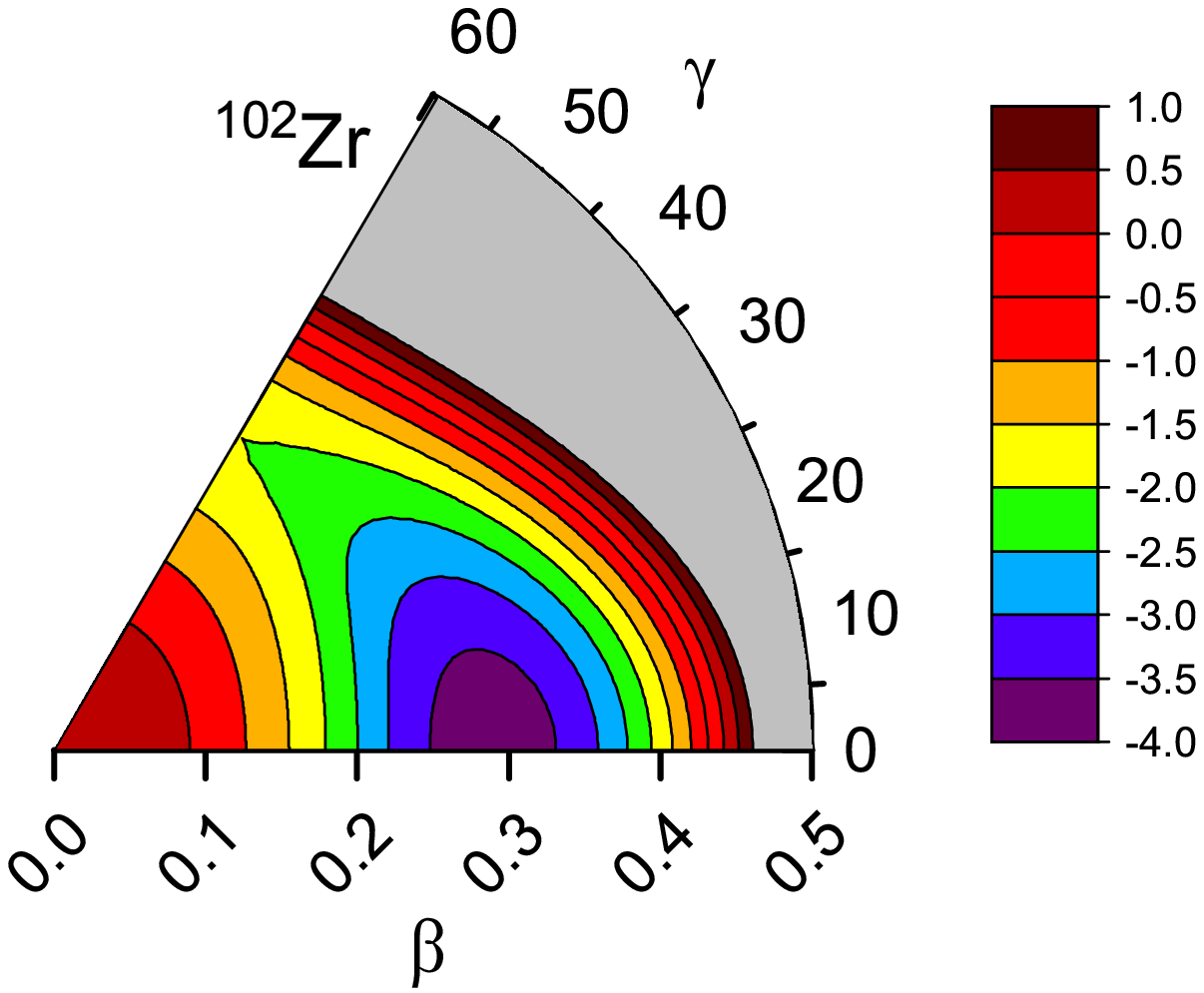} \\
\includegraphics[scale=0.2]{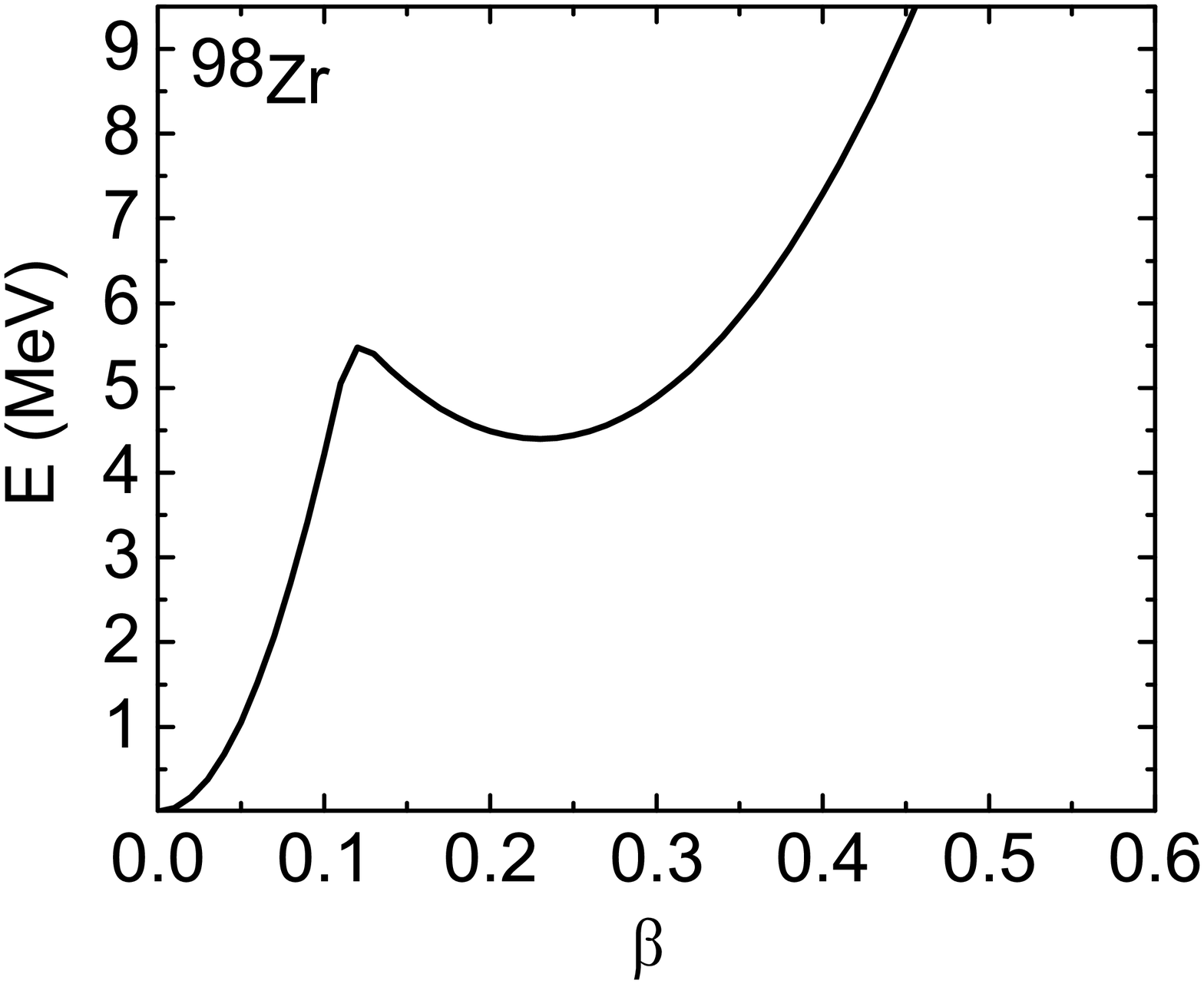}&
\includegraphics[scale=0.2]{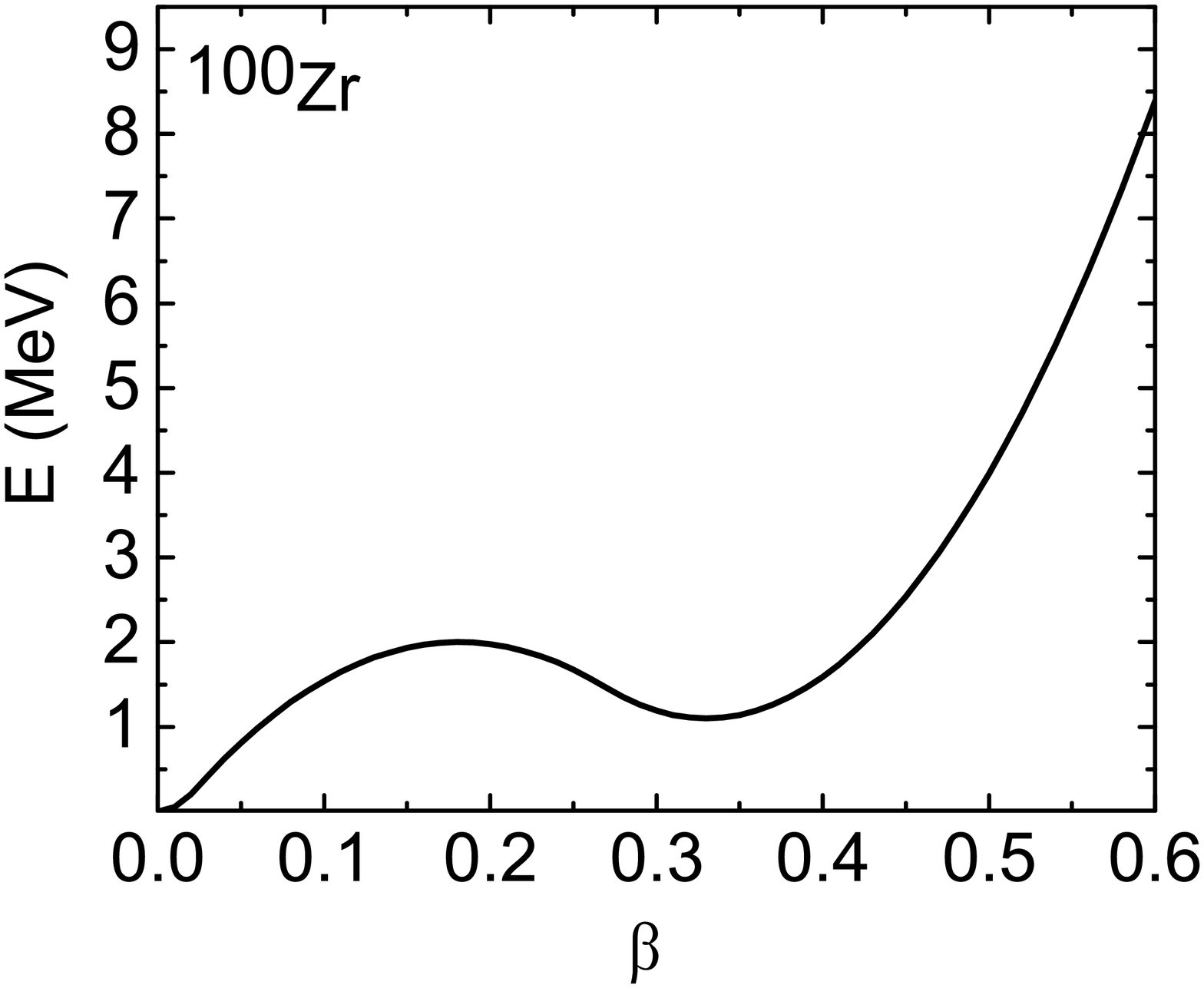}&
\includegraphics[scale=0.2]{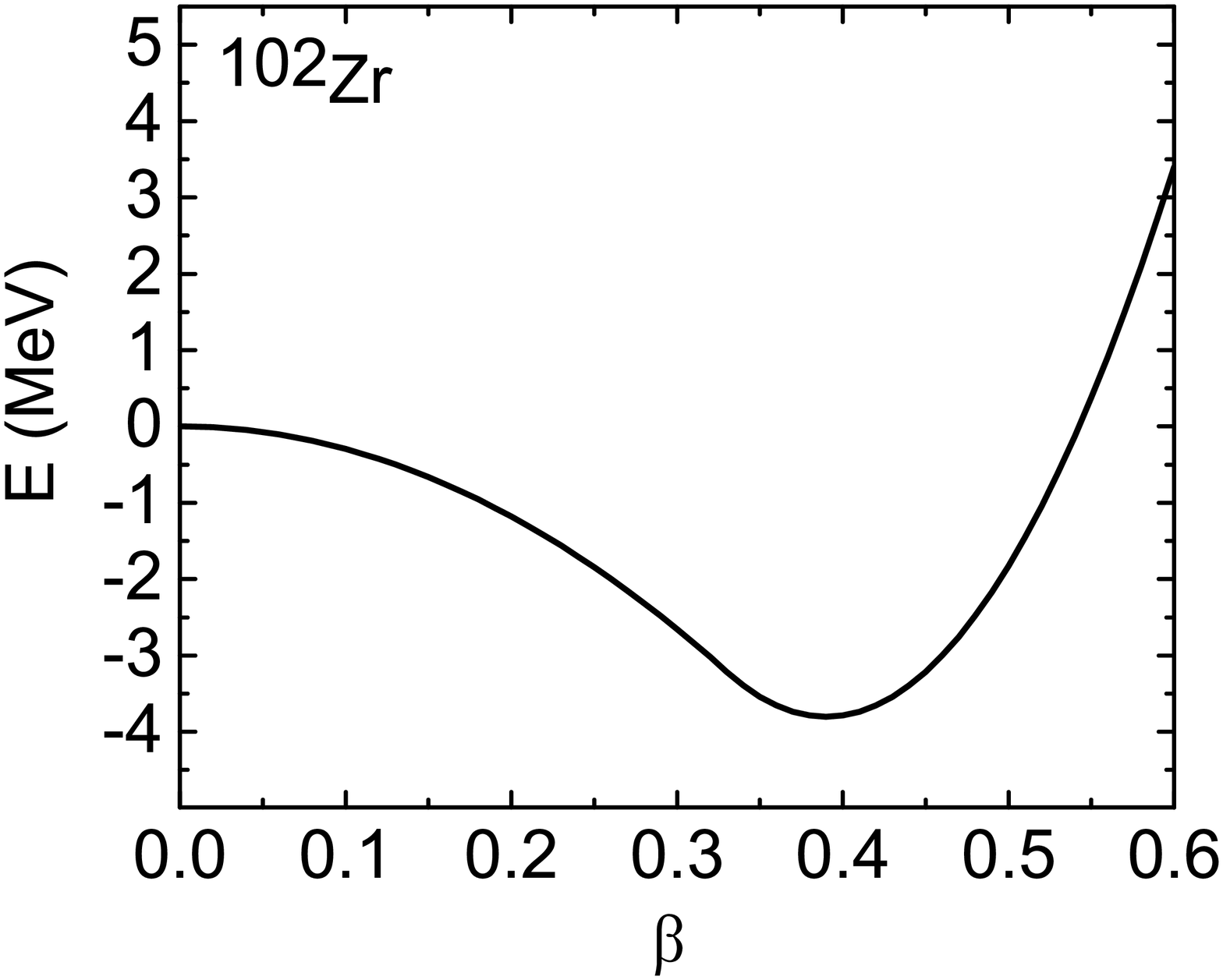}
\end{tabular}
\caption{\label{Fig5}Collective potential for the even-even Zr isotopes obtained by fitting the experimental data for the low-lying collective quadrupole states}
\end{figure*}
\subsection{Potential}

The resulting potentials $V(\beta,\gamma)$ for all considered Zr isotopes are presented in Fig.\ref{Fig5}, where the value of $C_\gamma$ is taken to be positive. However, the results of calculations of the excitation energies and transition probabilities will not change if we  take the negative value of $C_\gamma$ keeping its absolute value unchanged. In both cases the potential support the axial symmetry and this is a limitation of the present consideration. The absolute value of $C_\gamma$ is fixed to be equal to 50 MeV in order to reproduce a reasonable value of the frequency of the $\gamma$-vibrations close to 1.5 MeV.

The $\beta$-dependence of the collective potential taken for  $\gamma=0$ is illustrated in Fig.\ref{Fig5}. It is seen that in $^{94,96,98,100}$Zr spherical and deformed minima coexist. In $^{92}$Zr the deformed minimum is only intended. In $^{94}$Zr deformed minimum is very shallow,  in  $^{96,98,100}$Zr it is clearly isolated, and in  $^{102}$Zr there is only deformed minimum.

It is seen in Fig.\ref{Fig5} that the spherical minimum of the potential in $^{96,98}$Zr is more rigid than in lighter Zr isotopes. This fact correlate with the higher energies of the $2^+_1$ states in $^{96,98}$Zr
compare to $^{92,94}$Zr. This result is in accordance with the single particle level scheme of Zr
isotopes given in \cite{Taqi}. According to this scheme, the lowest neutron single particle level in the shell above $N=50$ is $d_{5/2}$. The following single particle level $s_{1/2}$ is located above $d_{5/2}$
by 1033 keV, and the single particle level $g_{7/2}$ is separated from $s_{1/2}$ by 1940 keV. Thus,
single particle neutron states $d_{5/2}$ and $s_{1/2}$ play the role of subshells and this explains a larger rigidity of the potential near the spherical minimum in $^{96,98}$Zr,
 in contrast to the lighter Zr isotopes.

\section{Conclusion}

We have studied a possibility to describe the properties of the low-lying collective quadrupole states of $^{92-102}$Zr based on the five dimensional Bohr collective Hamiltonian. Both $\beta$ and $\gamma$ shape collective variables are included into consideration. The  $\beta$-dependence of the potential energy is fixed to describe the experimental data in a best possible way. However, the $\gamma$-dependence of the potential is introduced in a simple way favoring axial symmetry at large
$\beta$. The resulting potential evolve with $A$ increase from having only one spherical minimum, through the potentials having both spherical and deformed minima, to the potential with one deformed minimum in $^{102}$Zr.

A detailed comparison of the energy spectra, the $E2$ transition probabilities and a discussion of the characteristics  of the resulting wave functions was carried out.  A $\beta$-dependence of the wave functions is presented in a set of figures illustrating their distribution over $\beta$ at $\gamma =0$.


\section{Acknowledgments}
The authors express their gratitude to the RFBR (grant 20--02-00176) and to the Heisenberg--Landau Program for support. One of the author, T.M.S., acknowledges support from Russian Government Subsidy Program of the Competitive Growth of Kazan Federal University.

\end{document}